\documentclass[rmp,aps,twocolumn]{revtex4}

\usepackage{graphics}
\usepackage{epsfig}

\newcommand{\be}{\begin{equation}}
\newcommand{\ee}{\end{equation}}
\newcommand{\ba}{\begin{eqnarray}}
\newcommand{\ea}{\end{eqnarray}}
\newcommand{\ban}{\begin{eqnarray*}}
\newcommand{\ean}{\end{eqnarray*}}
\newcommand{\braket}[2]{\mbox{$ \langle #1 | #2 \rangle $}}
\newcommand{\sandwich}[3]{\mbox{$ \langle #1 | #2 | #3 \rangle $}}
\newcommand{\ket}[1]{\mbox{$ | #1 \rangle $}}
\newcommand{\bra}[1]{\mbox{$ \langle #1 | $}}

\newcommand{\hilb}{{\cal{H}}}

\newcommand{\si}{\sigma}

\newcommand{\demi}{\frac{1}{2}}
\newcommand{\compl}{\begin{picture}(8,8)\put(0,0){C}\put(3,0.3){\line(0,1){7}}\end{picture}}
\newcommand{\one}{\leavevmode\hbox{\small1\normalsize\kern-.33em1}}

\newcommand{\moy}[1]{\langle #1 \rangle}

\newcommand{\bed}{\[}
\newcommand{\eed}{\]}
\newcommand{\beq}{\begin{equation}}
\newcommand{\eeq}{\end{equation}}
\newcommand{\beqa}{\begin{eqnarray}}
\newcommand{\eeqa}{\end{eqnarray}}
\newcommand{\tr}{\mathop{\mathrm{tr}}}

\begin{document}
\title{Quantum cloning}

\author{Valerio Scarani}
\email{valerio.scarani@physics.unige.ch} \affiliation{Group of
Applied Physics, University of Geneva, 20, rue de
l'Ecole-de-M\'edecine, CH-1211 Geneva 4, Switzerland}
\author{Sofyan Iblisdir}
\email{sofyan.iblisdir@physics.unige.ch} \affiliation{Group of
Applied Physics, University of Geneva, 20, rue de
l'Ecole-de-M\'edecine, CH-1211 Geneva 4, Switzerland}
\author{Nicolas Gisin}
\email{nicolas.gisin@physics.unige.ch} \affiliation{Group of
Applied Physics, University of Geneva, 20, rue de
l'Ecole-de-M\'edecine, CH-1211 Geneva 4, Switzerland}
\author{Antonio Ac\'{\i}n} \email{antonio.acin@icfo.es}
\affiliation{ICFO-Institut de Ci\`encies Fot\`oniques, 29 Jordi
Girona, E-08034 Barcelona, Spain}

\begin{abstract}
The impossibility of perfectly copying (or cloning) an arbitrary
quantum state is one of the basic rules governing the physics of
quantum systems. The processes that perform the optimal
approximate cloning have been found in many cases. These "quantum
cloning machines" are important tools for studying a wide variety
of tasks, e.g.~state estimation and eavesdropping on quantum
cryptography. This paper provides a comprehensive review of
quantum cloning machines (both for discrete-dimensional and for
continuous-variable quantum systems); in addition, it presents the
role of cloning in quantum cryptography, the link between optimal
cloning and light amplification via stimulated emission, and the
experimental demonstrations of optimal quantum cloning.
\end{abstract}

\maketitle

\tableofcontents

\newpage
\section{Cloning of quantum information}

\subsection{Introduction}

The concept of "information" is shaping our world: communication,
economy, sociology, statistics... all benefit from this
wide-encompassing notion. During the last decade or so,
information entered physics from all sides: from cosmology (e.g.
entropy of black-holes\footnote{The widely discussed topic of
black-hole evaporation is also a matter of information: is all the
information that has entered a black-hole lost forever ---
technically, does irreversible non-unitary dynamics exist in
nature?}) to quantum physics (the entire field of quantum
information processing). Some physicists even try to reduce all
natural sciences to mere information \cite{fuc02,brz02,col04}. In
this review, we concentrate on one of the essential features of
information: {\em the possibility to copy it}. One might think
that this possibility is an essential feature of any "good"
encoding of information. This is however not the case: when
information is encoded in quantum systems, in general it cannot be
replicated without introducing errors. This limitation, however,
does not make quantum information useless --- quite the contrary,
as we are going to show.

But we should first answer a natural question: why should one
encode information in quantum systems? Well, in the final
analysis, the {\em carriers of information} can only be physical
systems ("information is physical", as Rolf Landauer summarized
it); and ultimately, physical systems obey the laws of quantum
physics. So in some sense, the question that opened this paragraph
can be answered with another question: do you know any carriers of
information, other than quantum systems? The answer that most
physicists give is, "No, because everything is quantum" ---
indeed, the boundary between the classical and the quantum world,
if any such boundary exists, has not been identified yet. Other
reasons to be interested in quantum information will soon become
clear.

Still, even if the carrier of information is a quantum system, its
encoding may be classical. The most striking example found in
nature is DNA: information is encoded by molecules, which are
definitely quantum systems; but it is encoded in the "nature" of
the molecules (adenine, thymine, cytosine, guanine), not in their
state\footnote{This does not necessarily imply that the way Nature
processes this information is entirely classical: this point is an
open question.}. Such an encoding is classical, because one cannot
find a superposition of "being adenine" and "being thymine". If
information is encoded this way, it can be replicated perfectly:
this process is called {\em cloning}, nature performs it and
biologists are struggling to master it as well.

Here, we concentrate on the {\em quantum encoding of information},
when information is encoded in the {\em state $\psi$ of quantum
systems}. The process of replicating the state, written
$\psi\rightarrow \psi\otimes\psi$ and called cloning as well, can
be done perfectly and with probability 1 if and only if a basis to
which $\psi$ belongs is known. Otherwise, perfect cloning is
impossible: either the copies are not perfect, or they are perfect
but sometimes the copying process simply gives no outcome. These
are the content and the consequences of the {\em no-cloning
theorem} of quantum information. Similar to Heisenberg's
uncertainty relations, the no-cloning theorem defines an intrinsic
impossibility, not just a limitation of laboratory physics.

After some thinking though, one may object that the possibility of
classical telecommunication contradicts the no-cloning theorem:
after all, information travelling in optical fibers is encoded in
the state of light, so it should be a quantum encoding; and this
information is amplified several times from the source to the
receiver, so it should degrade. Indeed, it does. However, a
telecom signal consists of a large number of photons prepared in
the very same quantum state; so, amplification in telecom amounts
to producing some new copies of $\psi$ out of $\psi^{\otimes N}$.
In short, the no-cloning theorem {\em does} apply to the
amplification of telecom signals, because spontaneous emission is
always present in amplifiers; but the copy is almost perfect,
because stimulated emission is the dominating effect. The
sensitivity of present-day devices is such that the quantum limit
should be reached in the foreseeable future\footnote{The security
parameter for the acceptable error is presently set at
$e=10^{-9}$. Let's make a simple estimate of the ultimate quantum
limit that corresponds to it: the signal is a coherent state
$\ket{\alpha}$, and let's say that an error is possible for the
vacuum component, because for that component there is no
stimulated emission. Then $e\approx
|\braket{\alpha}{0}|^2=\exp(-|\alpha|^2)$, which is equal to
$10^{-9}$ for an average number of photons $|\alpha|^2\approx 20$.
In actual networks, a telecom pulse that has travelled down a
fiber reaches the amplifier with an intensity of some 100 photons
on average.}.

We left for the end of the introduction the most surprising: an
encoding of information that obeys the no-cloning theorem is {\em
helpful}. The impossibility of perfectly copying quantum
information does not invalidate the entire concept of quantum
information. Quite the opposite, it provides an illustration of
its power. There is no way for someone to perfectly copy the state
of a quantum system, for a clever encoding of information which
uses a set of non-orthogonal states. Consequently, if such a
system arrives unperturbed at a receiver, then, for sure, it has
not been copied by any adversary. Hence, due to the no-cloning
theorem, quantum information provides a means to perform some
tasks that would be impossible using only ordinary information,
such as detecting any eavesdropper on a communication channel:
this is the idea of quantum cryptography.

The outline of the review will be given in paragraph
\ref{sssplan}, after some concepts have been introduced. We start
by stating and demonstrating the no-cloning theorem, and by
sketching its history.

\subsection{The no-cloning theorem}
\label{ssthm}

It is well-known that one cannot measure the state $\ket{\psi}$ of
a single quantum system: the result of any single measurement of
an observable $A$ is one of its eigenstates, bearing only very
poor information about $\ket{\psi}$, namely that it must not be
orthogonal to the measured eigenstate. To reconstruct $\ket{\psi}$
(or more generally, any mixed state $\rho$) one has to measure the
average values of several observables, and this implies making
statistical averages over a large number of identically prepared
systems \cite{woo89}. One can imagine how to circumvent this
impossibility in the following way: take the system in the unknown
state $\ket{\psi}$ and let it interact with $N$ other systems
previously prepared in a blank reference state $\ket{R}$, in order
to obtain $N+1$ copies of the initial state: \ba
\ket{\psi}\otimes\ket{R}...\otimes\ket{R}
&\stackrel{?}{\longrightarrow}
\ket{\psi}\otimes\ket{\psi}...\otimes\ket{\psi}\,. \label{ideal}
\ea Such a procedure would allow one to determine the quantum
state of a single system, without even measuring it because one
could measure the $N$ new copies and leave the original untouched.
The no-cloning theorem of quantum information formalizes the
suspicion that such a procedure is impossible:

{\em No-cloning theorem: No quantum operation exists that can
duplicate perfectly an arbitrary quantum state.}

The theorem can be proved with a {\em reductio ad absurdum} by
considering the $1\rightarrow 2$ cloning. The most general
evolution of a quantum system is a trace-preserving
completely-positive (CP) map. A well-known theorem \cite{kra83}
says that any such map can be implemented by appending an
auxiliary system (ancilla) to the system under study, let the
whole undergo a unitary evolution, then trace out the ancilla. So
let us suppose that perfect cloning can be realized as a unitary
evolution, possibly involving an ancilla (the "machine"): \ba
\ket{\psi}\otimes\ket{R}\otimes\ket{{\cal M}}
&\stackrel{U?}{\longrightarrow}
\ket{\psi}\otimes\ket{\psi}\otimes\ket{{\cal M}(\psi)}\,.
\label{perf2}\ea In particular then, for two orthogonal states
labelled $\ket{0}$ and $\ket{1}$, we have: \ban
\ket{0}\otimes\ket{R}\otimes\ket{{\cal M}}
&\longrightarrow& \ket{0}\otimes\ket{0}\otimes\ket{{\cal M}(0)}\,,\\
\ket{1}\otimes\ket{R}\otimes\ket{{\cal M}} &\longrightarrow&
\ket{1}\otimes\ket{1}\otimes\ket{{\cal M}(1)}\,. \ean But because
of linearity (we omit tensor products) these conditions imply:
\ban \big(\ket{0}+\ket{1}\big)\ket{R}\,\ket{{\cal M}}
&\longrightarrow& \ket{00}\,\ket{{\cal M}(0)}+
\ket{11}\,\ket{{\cal M}(1)}\,.\ean The r.h.s. cannot be equal to
$\big(\ket{0}+\ket{1}\big) \big(\ket{0}+\ket{1}\big)\,\ket{{\cal
M}(0+1)}=
\big(\ket{00}+\ket{10}+\ket{01}+\ket{11}\big)\,\ket{{\cal
M}(0+1)}$. So (\ref{perf2}) may hold for states of an orthonormal
basis, but cannot hold for all states. This concludes the proof
using only the linearity of quantum transformations following the
work of \textcite{woo82}; a slightly different proof, using more
explicitly the properties of unitary operations, can be found in
Sect.~9-4 of Peres' textbook \cite{per95}.

\subsection{History of the no-cloning theorem}
\label{sechist}

\subsubsection{When "wild" ideas trigger deep results}

\begin{center}
\begin{figure}
\includegraphics[width=8cm]{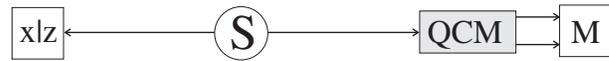}
\caption{Setup devised by Herbert to achieve signaling. A source S
produces pairs of qubits in a maximally entangled state; on the
left, Alice measures either $\si_x$ or $\si_z$. On the right, Bob
applies a perfect quantum cloning machine (QCM) and then measures
the two clones (the measurement M may be individual or
collective). } \label{figsignaling}
\end{figure}
\end{center}

Historically, the no-cloning theorem did not spring out of deep
thoughts on the quantum theory of measurement. The triggering
event was a rather unconventional proposal by Nick Herbert to use
quantum correlations to communicate faster-than-light
\cite{her82}. Herbert called his proposal FLASH, as an acronym for
"First Light Amplification Superluminal Hookup". The argument goes
as follows (Fig.~\ref{figsignaling}). Consider two parties, Alice
and Bob, at an arbitrary distance, sharing two qubits\footnote{A
qubit is a two-dimensional quantum system. In this paper, the
mathematics of qubits are used extensively: we use the standard
notations of quantum information, summarized in Appendix
\ref{appa} together with some useful formulae. We shall also use
the term {\em qudit} to designate a $d$-level quantum system.} in
the singlet state
$\ket{\Psi^-}=\frac{1}{\sqrt{2}}\left(\ket{0}_A\otimes\ket{1}_B -
\ket{1}_A\otimes\ket{0}_B\right)$. On her qubit, Alice measures
either $\si_x$ or $\si_z$. Because of the properties of the
singlet, if Alice measures $\si_z$, she finds the eigenstate
$\ket{0}$ (resp. $\ket{1}$) with probability $\demi$, and in this
case she prepares Bob's qubit in the state $\ket{1}$ (resp.
$\ket{0}$). Without any knowledge on Alice, Bob sees the mixed
state $\demi\ket{0}\bra{0}+\demi\ket{1}\bra{1}=\demi\one$, just as
if Alice had done nothing. Similarly, if Alice measures $\si_x$,
she finds the eigenstate $\ket{+}$ (resp. $\ket{-}$) with
probability $\demi$, and in this case she prepares Bob's qubit in
the state $\ket{-}$ (resp. $\ket{+}$). Again, without any
knowledge on Alice, Bob sees the mixed state
$\demi\ket{+}\bra{+}+\demi\ket{-}\bra{-}=\demi\one$.

However, suppose that Bob has a perfect $1\rightarrow 2$ cloner,
QCM in Fig.~\ref{figsignaling}, and that he has his qubit pass
through it. Now, if Alice measures $\si_x$, Bob's mixture is
$\rho_x \,=\,\demi \ket{++}\bra{++} + \demi \ket{--}\bra{--}$; if
Alice measures $\si_z$, Bob's mixture is $\rho_z\,=\,\demi
\ket{00}\bra{00} + \demi \ket{11}\bra{11}$. It is easily verified
that $\rho_x\neq \rho_z$ (for instance,
$\sandwich{01}{\rho_x}{01}=\frac{1}{4}$ while
$\sandwich{01}{\rho_z}{01}=0$). Thus, at least with some
probability, by measuring his two perfect clones, Bob could know
the measurement that Alice has chosen without any communication
with her.

This is an obvious violation of the no-signaling condition, but
the argument was clever --- that is why it was published
\cite{per02} --- and triggered the responses\footnote{NG: "I
vividly remember the conference held somewhere in Italy for the
90th birthday of Louis de Broglie. I was a young PhD student.
People around me were all talking about a "FLASH communication"
scheme, faster than light, based on entanglement. This is where -
I believe - the need for a no-cloning theorem appeared. Zurek and
Milonni were among the participants."} of \textcite{woo82},
\textcite{die82}, \textcite{mil82} and slightly later
\textcite{man83}. In these papers, the no-cloning theorem was
firmly established as a consequence of the linearity of quantum
mechanics. It was also shown that the best-known amplification
process, spontaneous and stimulated emission of a photon by an
excited system, was perfectly consistent with this no-go theorem
\cite{woo82,mil82,man83}.

\subsubsection{Missed opportunities}

Once the simplicity of the no-cloning theorem is noticed, one
cannot but wonder why its discovery was delayed until 1982. There
is no obvious answer to this question. But we can review two
"missed opportunities".

In 1957, during a sabbatical in Japan, Charles Townes worked out
with Shimoda and Takahasi the phenomenological equations which
describe the amplification in the maser that he had demonstrated
four years before \cite{shi57,tow02}. In this paper, see
discussion in paragraph \ref{sssfiber} for more details, some rate
equations appear from which the "fidelity" of optimal quantum
cloning processes\footnote{Specifically, universal symmetric
$N\rightarrow M$ cloning of qubits, see below.} immediately
follows. At that time however, nobody used to look at physics in
terms of information; so in particular, nobody thought of
quantifying amplification processes in terms of the accuracy to
which the input state is replicated.

The second missed opportunity involved Eugene Paul Wigner. In a
{\em Festschrift}, he tackled the question of biological cloning
\cite{wig61}. Wigner tentatively identified the "living state"
with a pure quantum mechanical state, noted $\nu$; he then noticed
that, among all the possible unitary transformations, those that
implement $\nu\otimes w\rightarrow \nu\otimes\nu\otimes r$ are a
negligible set --- but he did not notice that {\em no
transformation} realizes that task for any $\nu$, which would have
been the no-cloning theorem. From his observation, Wigner
concluded that biological reproduction "appears to be a miracle
from the point of view of the physicist". We know nowadays that
his tentative description of the living state is not correct, and
that reproduction is possible because the encoding in DNA is
classical (see the Introduction of this review).

\subsubsection{From no-cloning to optimal cloning}

Immediately after its formulation, the no-cloning theorem became
an important piece of physics, cited in connection with both
no-signaling \cite{ghi83,bus87} and amplification \cite{yue86}.
Interestingly, no-cloning was invoked to argue for the security of
quantum cryptography from the very beginning \cite{ben84}. Sect.
9-4 of Peres' book \cite{per95} is a good review of the role of
the theorem before 1996. In the first months of that very year,
\textcite{bar96} considered the possibility of the perfect cloning
of non-commuting mixed states, and reach the same no-go conclusion
as for pure states. Everything fell into place.

The situation suddenly changed a few months later: in the
September 1996 issue of Physical Review A, Vladim\'{\i}r Bu\v{z}ek
and Mark Hillery published a paper whose title is "Quantum
copying: beyond the no-cloning theorem" \cite{buz96}. Of course,
they did not claim that the no-go theorem was wrong. But the
theorem applied only to {\em perfect} cloning, whereas Bu\v{z}ek
and Hillery suggested the possibility of {\em imperfect} cloning.
Specifically (see \ref{ssbh} below for all details), they found a
unitary operation \ba
\ket{\psi}_A\otimes\ket{R}_B\otimes\ket{{\cal M}}_M
&\longrightarrow \ket{\Psi}_{ABM}\ea such that the partial traces
on the original qubit A and on the cloned qubit B satisfy \ba
\rho_A\,=\,\rho_B&=&F\ket{\psi}\bra{\psi}+(1-F)
\ket{\psi^{\perp}}\bra{\psi^{\perp}} \ea with a fidelity $F$ that
is "not too bad" ($\frac{5}{6}$) and is the same for any input
state $\ket{\psi}$. The Bu\v{z}ek-Hillery unitary transformation
is the first {\em Quantum Cloning Machine}; it triggered an
explosion in the number of investigations on quantum cloning.

\subsection{Quantum Cloning Machines (QCM): generalities}

\subsubsection{Definition of cloning}

Any interaction (i.e. any CP map) between two quantum systems A
and B, possibly mediated by an ancilla M, has the effect of
"shuffling" the quantum information between all the sub-systems.
When the input state takes the form $\ket{\psi}_A\ket{R}_B$, then
at the output of any CP map the quantum information contained in
$\ket{\psi}$ will have been somehow distributed among A and B (and
possibly the ancilla). This suggests the following definition of
the process of {\em cloning of pure states}, that we generalize
immediately to the case of $N\rightarrow M$ cloning: \ba
\left(\ket{\psi}^{\otimes N}\right)\,\otimes\,
\left(\ket{R}^{\otimes M-N}\right)\,\otimes\,\ket{{\cal M}}
&\stackrel{U}{\longrightarrow}& \ket{\Psi}
\label{generalcloning}\ea where $\ket{\psi}$ is the state of
${\cal H}$ to be copied, $\ket{R}$ is a reference state
arbitrarily chosen in the same Hilbert space ${\cal H}$, and
$\ket{{\cal M}}$ is the state of the ancilla. In other words:
\begin{itemize}
\item The fact that the process is a form of cloning is determined
by the form of the input state, l.h.s. of (\ref{generalcloning}):
$N$ particles ("originals") carry each the pure state $\ket{\psi}$
to be copied. In particular then, the $N$ originals are {\em
disentangled} from the $M-N$ particles that are going to carry the
"copies" (that start in a blank state) and from the "ancillae". In
fact, sometimes (e.g. in paragraphs \ref{ssasym}, \ref{ssstate})
it will be convenient to consider that the copies and the ancillae
start in an entangled state. This does not contradict
(\ref{generalcloning}): one simply omits to mention a "trivial"
part of the QCM, that prepares the copies and the ancillae in the
suitable state. It is important to stress that {\em we consider
only pure states as inputs}; to our knowledge, there are no
results on QCM that would be optimal for mixed states.

\item The cloning process (\ref{generalcloning}) is defined by the
Quantum Cloning Machine (QCM), which is the trace-preserving CP
map, or equivalently the pair \ba \mbox{QCM} &=& \big\{U,
\ket{{\cal M}}\big\}\,. \label{qcm}\ea A QCM can be seen as a
"quantum processor" $U$, that processes the input data according
to some "program" $\ket{{\cal M}}$. Examples of QCM, that produce
clones of very different quality, are easily found: just take the
identity $\ket{\psi}_A\ket{R}_B\rightarrow \ket{\psi}_A\ket{R}_B$
that transfers no information from A to B, or the swap
$\ket{\psi}_A\ket{R}_B\rightarrow \ket{R}_A\ket{\psi}_B$ that
transfers all the information from A to B --- both unitary
operations, with no ancilla. The possibility to define coherent
combinations of such processes suggests that non-trivial QCM can
be found: and indeed, we'll see that this intuition is basically
correct (see particularly \ref{ssasym}), although not fully,
because ancillae play a crucial role.

\end{itemize}

\subsubsection{Fidelity, and the Glossary of QCM}

Having defined the meaning of "cloning", we can introduce the
basic glossary that is used in the study and classification of
QCMs. The very first object to define is a {\em figure of merit}
according to which the output of the QCM should be evaluated. The
usual figure of merit is the {\em single-copy fidelity}, called
simply {\em fidelity} unless some ambiguity is possible. This is
defined for each of the outputs $j=1...M$ of the cloning machine,
as the overlap between $\rho_j$ and the initial state
$\ket{\psi}$: \ba F_{j}&=&
\sandwich{\psi}{\rho_j}{\psi}\,,\;\;j=1,...,M\label{fiddef}\ea
where $\rho_j$ is the partial state of clone $j$ in the state
$\ket{\Psi}$ defined in (\ref{generalcloning}). Note that the
worst possible fidelity for the cloning of a $d$-dimensional
quantum system is $F_j=\frac{1}{d}$, obtained if $\rho_j$ is the
maximally mixed state $\frac{\one}{d}$.

The following, standard classification of QCM follows:
\begin{itemize}

\item A QCM is called {\bf universal} if it copies equally well
all the states, that is, if $F_j$ is independent of $\ket{\psi}$.
The notation UQCM is often used. Non-universal QCM are called {\em
state-dependent}.

\item A QCM is called {\bf symmetric} if at the output all the
clones have the same fidelity, that is if $F_j=F_{j'}$ for all
$j,j'=1,...,M$. For asymmetric QCM, further classifications are
normally needed: for instance, in the study of $1\rightarrow 3$
asymmetric QCM, one may restrict to the case $F_1\neq F_2=F_3$
(that we shall write $1\rightarrow 1+2$) or consider the general
case $1\rightarrow 1+1+1$ where all the three fidelities can be
different.

\item A QCM is called {\bf optimal} if, for a given fidelity of
the original(s), the fidelities of the clones are the maximal ones
allowed by quantum mechanics. More specifically, if ${\cal S}$ is
the set of states to be cloned, optimality can be defined by
maximizing either the average fidelity over the states
$\bar{F}=\int_{\cal S} d\psi F(\psi)$, or the minimal fidelity
over the states $F_{min}=\min_{\psi\in{\cal S}}F(\psi)$. These
definitions often coincide.

\end{itemize}
According to this classification, for instance, the
Bu\v{z}ek-Hillery QCM is the optimal symmetric UQCM for the
cloning $1\rightarrow 2$ of qubits. The generalization to optimal
symmetric UQCM for the cloning $N\rightarrow M$ have been rapidly
found, first for qubits \cite{gis97,bru98a}, then for
arbitrary-dimensional systems \cite{wer98,key99}. Also the family
of optimal asymmetric UQCM for the cloning $1\rightarrow 1+1$ of
arbitrary dimension has been fully characterized
\cite{cer00b,bra01,ibl04,ibl05,fiu05}. The study of the optimal
universal asymmetric QCM $N\rightarrow M_1+M_2$ has been
undertaken later, motivated by the fact that the $2\rightarrow
2+1$ QCM is needed for the security analysis of quantum
cryptography protocols. As the reader may easily imagine, the full
zoology of QCM has not been explored: there are hard problems that
wait for a motivation. For instance, very few examples of optimal
state-dependent QCM are known.

\subsubsection{Outline of the paper}
\label{sssplan}

The outline of this review is as follows. In Section \ref{sec2},
we review the cloning of discrete quantum systems, presenting the
QCMs for qubits and stating the generalizations to
larger-dimensional systems. We introduce at the end of this
section the link between cloning and state estimation. In Section
\ref{sec3}, we will review the cloning of continuous variables.
Section \ref{sec4} is devoted to the application of quantum
cloning for eavesdropping in quantum cryptography. The last two
sections are devoted to the realization of quantum cloning.
Section \ref{sec5} shows how the amplification based on the
interplay of spontaneous and stimulated emission achieves optimal
cloning of discrete systems encoded in different modes of the
light field. We present a self-contained derivation of this claim.
In Section \ref{sec6}, we review the experimental proposals and
demonstrations of cloning, for the polarization of photons and for
other physical systems.

Some topics related to cloning are omitted in this review. One of
them is {\em probabilistic exact cloning} \cite{dua98,pat99}:
while in the spirit of Bu\v{z}ek-Hillery one circumvents the
no-cloning theorem by allowing imperfect cloning, in probabilistic
cloning one wants to always obtain a perfect copy, but the price
is that the procedure works only with some probability. This is
related to unambiguous state discrimination procedures in
state-estimation theory. In comparison to probabilistic cloning,
the cloning procedures {\em \`a la} Bu\v{z}ek-Hillery that we
describe in this review are called {\em deterministic cloning},
because the desired result, namely imperfect copying, is always
obtained. Hybrid strategies between probabilistic-exact and
deterministic-imperfect cloning have also been studied and
compared to results of state-estimation theory \cite{che99}.

Another topic that will be omitted is {\em telecloning}, that is,
cloning at a distance. In this protocol, a party Alice has a copy
of an unknown quantum state, and wants to send the best possibly
copy to each of $M$ partners. An obvious procedure consists in
performing locally the optimal $1\rightarrow M$ cloning, then
teleporting the $M$ particles to each partner; this strategy
requires $M$ singlets (that is, $M$ bits of entanglement or
e-bits) and the communication of $2M$ classical bits. It has been
proved \cite{mur99} that other strategies exist, that are much
cheaper in terms of the required resources. In particular, the
partners can share a suitable entangled state of only $O(\log_2M)$
e-bits, and in this case the classical communication is also
reduced to public broadcasting of two bits.

\subsection{No-cloning and other "limitations"}

As a last general discussion, we want to briefly sketch the link
between cloning and other "limitations" that are found in quantum
physics; specifically, the no-signaling condition and the
uncertainty relations.

\subsubsection{Relation to no-signaling} \label{secsign}

As we said in paragraph \ref{sechist}, a perfect cloner would
allow signaling through entanglement alone. Shortly after the idea
of imperfect cloning was put forward, \cite{gis98} noticed that
one can also study optimal imperfect cloning starting from the
requirement that no-signaling should hold. The proof was given for
universal symmetric $1\rightarrow 2$ cloning for qubits. The idea
is to require that the input state
$\ket{\psi}\bra{\psi}=\demi\left(\one+\hat{m}\cdot\sigma\right)$
is copied into a two qubit state such that the two one-qubit
partial states are equal and read $\rho_1=\rho_2=\demi\left(
\one+\eta\hat{m}\cdot\sigma\right)$: i.e., the Bloch vector points
in the same direction as for the original but is shrunk by a
factor $\eta$ ({\em shrinking factor}), related to the fidelity
defined in (\ref{fiddef}) through $F=\frac{1+\eta}{2}$. On the one
hand, we know from the no-cloning theorem that $\eta=1$ is
impossible and would lead to signaling; on the other hand,
$\eta=0$ is obviously possible by simply throwing the state away
in a non-monitored mode and preparing a new state at random. So
there must be a largest shrinking factor $\eta$ compatible with
the no-signaling condition.

The form of the partial states $\rho_{1,2}$ implies that the state
of systems 1 and 2 after cloning should read \ba
\rho_{out}(\psi)&=&\frac{1}{4}\Big(\one_4\,+\,\eta\,\big(\hat{m}\cdot\sigma\otimes
\one + \one\otimes \hat{m}\cdot\sigma\big) \nonumber\\&& +\,
\sum_{i,j=x,y,z}t_{ij}\si_i\otimes\si_j\Big)\,. \ea The tensor
$t_{ij}$ has some structure because of the requirement of
universality, that implies covariance: \ba
\rho_{out}\big(U\ket{\psi}\big)&=&U\otimes U\,\rho_{out}(\psi)\,
U^{\dagger}\otimes U^{\dagger}\,. \ea This means that the
following two procedures are equivalent: to apply a unitary $U$ on
the original and then the cloner, or to apply the cloner first and
then $U$ to both copies.

These are the requirements of universality, $\eta$ accounting for
imperfect cloning. With this definition of QCM, one can run again
the {\em Gedankenexperiment} discussed in paragraph \ref{sechist}
(Fig.~\ref{figsignaling}). Bob's mixtures after cloning read now
$\rho_x\,=\,\demi\,\rho_{out}(+x) \,+\,\demi\,\rho_{out}(-x)$ and
$\rho_z\,=\,\demi\,\rho_{out}(+z) \,+\,\demi\,\rho_{out}(-z)$.
No-signaling requires $\rho_x=\rho_z$. By using the fact that
density matrices must be positive operators, one finds after some
calculation the bound $\eta\leq \frac{2}{3}$ ($F\leq \frac{5}{6}$)
for any universal symmetric $1\rightarrow 2$ QCM for qubits. This
analysis alone does not say whether this bound can be attained;
but we know it can: the Bu\v{z}ek-Hillery QCM reaches up to it.
Thus, the no-signaling condition provides a bound for the fidelity
of quantum cloning, and this bound is tight since there exists a
QCM that saturates it; in turn, this provides a proof of the
optimality of the Bu\v{z}ek-Hillery QCM. The argument can be
generalized for the $1\rightarrow N$ symmetric cloning
\cite{sim01a}. Other QCMs on the edge of the no-signaling
condition have been described more recently \cite{nav03}.

In conclusion, the no-signaling condition has been found to
provide tight bounds for cloning --- in fact, this observation was
extended to any linear trace-preserving CP map \cite{sim01b}. The
converse statement also holds: no linear trace-preserving CP map
(so in particular, no QCM) can lead to signaling \cite{bru00}.
Finally, it has been proved recently that no-cloning is a feature
that holds for all non-local no-signaling theories \cite{mas05}.

\subsubsection{Relation to uncertainty relations and knowledge}

In addition to allowing signaling through entanglement alone,
perfect cloning would also violate one of the main tenets of
quantum mechanics, namely that the state of a single quantum
system cannot be known\footnote{Invoking the same argument as in
paragraph 9-4 of Peres' book \cite{per95}, perfect cloning would
thus lead to a violation of the second law of thermodynamics.
However, the cogency of this argument is disputed (see
\textcite{man05} for a recent analysis). In fact, to derive the
violation of the second law, one makes the assumptions that (i)
non-orthogonal states are deterministically distinguishable, and
(ii) entropies are computed using the quantum formalism. Clearly,
the two assumptions already look contradictory.}. If perfect
cloning were possible, one could know everything of a single
particle's state without even measuring it, just by producing
clones and measure these (see \ref{ssthm}). In turn, this would
invalidate quantum cryptography (see Section \ref{sec4}), and lead
to the violation of some information-theoretical principles, such
as Landauer's erasure principle \cite{ple01}. The link between
optimal cloning and the amount of knowledge that one can obtain on
the state of a limited number of quantum systems (in the limit,
just one) can be made quantitative, see \ref{ss4meas} in this
review.

Of course, perfect cloning would not invalidate the existence of
incompatible observables: having $N$ copies of an eigenstate of
$\si_z$ does not mean that the result of a measurement of $\si_x$
becomes deterministic. In particular, the relation between
observables $\Delta A\Delta B\geq \demi\left|\moy{[A,B]}\right|$
would still hold in the presence of perfect cloning.

\section{Cloning of discrete quantum systems}
\label{sec2}

In this Section, we review the main results about cloning of
discrete quantum systems, that is, systems described by the
Hilbert space ${\cal H}=\compl^d$. We start from the simplest
case, $1\rightarrow 2$ symmetric cloning for qubits, and describe
the Bu\v{z}ek-Hillery QCM (\ref{ssbh}). Then, we present the two
natural extensions: $N\rightarrow M$ symmetric cloning
(\ref{sssymnm}), and $1\rightarrow 1+1$ asymmetric cloning
(\ref{ssasym}). The last paragraph of this Section is devoted to
state-dependent cloning (\ref{ssstate}). An important remark when
comparing with the original articles: in this review, we use $d$
systematically for the dimension, and capital letters such as $N$
and $M$ for the number of quantum systems. This notation is
nowadays standard; however, until recently, $N$ was often used to
denote the dimension of the Hilbert space.

\subsection{Symmetric $1\rightarrow 2$ UQCM for qubits}
\label{ssbh}

\subsubsection{Trivial cloning}
\label{ssstriv}

In order to appreciate the performance of the optimal
$1\rightarrow 2$ cloning for qubits (Bu\v{z}ek-Hillery), it is
convenient to begin by presenting two trivial cloning strategies.
The {\em first trivial cloning strategy} is the
"measurement-based" procedure: one measures the qubit in a
randomly chosen basis and produces two copies of the state
corresponding to the outcome. Suppose that the original state is
$\ket{+ \vec{a}}$, whose projector is
$\demi(\one+\vec{a}\cdot\vec{\si})$, and that the measurement
basis are the eigenstates of $\vec{b}\cdot\vec{\si}$. With
probability $P_{\pm}=\demi(1\pm\vec{a}\cdot\vec{b})$, two copies
of $\ket{\pm \vec{b}}$ are produced; in either case, the fidelity
is $F_{\pm}=|\braket{+ \vec{a}}{\pm \vec{b}}|^2=P_{\pm}$. The
average fidelity is \ba F_{triv,1}&=& \int_{S_2}d\vec{b}
\left(P_{+}F_{+}+ P_{-}F_{-}\right) \nonumber\\ &=&\demi +\demi
\int_{S_2}d\vec{b}\left(
\vec{a}\cdot\vec{b}\right)^2\,=\,\frac{2}{3}\ea where $S_2$ is the
2-sphere of unit radius (surface of the Bloch sphere). This
cloning strategy is indeed universal: the fidelity is independent
of the original state $\ket{+ \vec{a}}$.

The {\em second trivial cloning strategy} can be called "trivial
amplification": let the original qubit fly unperturbed, and
produce a new qubit in a randomly chosen state. Suppose again that
the original state is $\ket{+ \vec{a}}$, and suppose that the new
qubit is prepared in the state $\ket{+ \vec{b}}$. We detect one
particle: the original one with probability $\demi$ and in this
case $F=1$; the new one with the same probability and in this case
the fidelity is $F=|\braket{+ \vec{a}}{+ \vec{b}}|^2=P_+$. Thus
the average single-copy fidelity is \ba F_{triv,2}&=&
\demi\,+\,\demi\, \int_{S_2}d\vec{b}\left(
\frac{1+\vec{a}\cdot\vec{b}}{2}\right)\,=\,\frac{3}{4}\,. \ea This
second trivial strategy is also universal. In conclusion, we shall
keep in mind that a fidelity of 75\% for universal $1\rightarrow
2$ cloning of qubits can be reached by a rather uninteresting
strategy.

\subsubsection{Optimal Symmetric UQCM (Bu\v{z}ek-Hillery)}

It's now time to present explicitly the symmetric UQCM for
$1\rightarrow 2$ cloning of qubits found by Bu\v{z}ek and Hillery
(B-H). This machine needs just one qubit as ancilla. Its action in
the computational basis of the original qubit is\footnote{We
rewrite, with a change of notation for the ancilla states,
Eq.~(3.29) of \textcite{buz96}.} \ba
\begin{array}{ccc} \ket{0}\ket{R}\ket{{\cal M}} &\rightarrow &
\sqrt{\frac{2}{3}}\, \ket{0}\ket{0}\ket{1}\,-\,
\sqrt{\frac{1}{3}}\,
\ket{\Psi^+}\ket{0}\\
\left(-\ket{1}\right)\ket{R}\ket{{\cal M}} &\rightarrow &
\sqrt{\frac{2}{3}}\, \ket{1}\ket{1}\ket{0}\,-\,
\sqrt{\frac{1}{6}}\,\ket{\Psi^+} \ket{1}\end{array}
\label{hbbase}\ea with
$\ket{\Psi^+}=\frac{1}{\sqrt{2}}\big[\ket{1}\ket{0}+\ket{0}\ket{1}\big]$.
By linearity, these two relations induce the following action on
the most general input state $\ket{\psi}=\alpha\ket{0}+
\beta\ket{1}$: \ba \ket{\psi}\ket{R}\ket{{\cal M}} &\rightarrow &
\sqrt{\frac{2}{3}}\,
\ket{\psi}\ket{\psi}\ket{\psi^{\perp}}\nonumber\\ &&\,-\,
\sqrt{\frac{1}{6}}\, \big[\ket{\psi}\ket{\psi^\perp}+
\ket{\psi^\perp}\ket{\psi}\big]\ket{\psi} \label{hbgen}\ea where
$\ket{\psi^\perp}=\alpha^{*}\ket{1}- \beta^*\ket{0}$.

From Eq.~(\ref{hbgen}), one sees immediately that A and B can be
exchanged, and in addition, that the transformation has the same
form for all input state $\ket{\psi}$. Thus, this QCM is symmetric
and universal. The partial states for the original and the copy
are \ba \rho_{A}=\rho_B&=&
\frac{5}{6}\ket{\psi}\bra{\psi}+\frac{1}{6}
\ket{\psi^\perp}\bra{\psi^\perp}\nonumber\\& =&
\frac{1}{2}\,\Big(\one + \frac{2}{3}\hat{m}\cdot\vec{\si}\Big)\,.
\label{rhoahb}\ea From the standpoint of both A and B then, the
B-H QCM shrinks the original Bloch vector $\hat{m}$ by a shrinking
factor $\eta=\frac{2}{3}$, without changing its direction. As
mentioned previously, the fidelity is
$F_A=F_B=\sandwich{\psi}{\rho_A}{\psi}=\frac{5}{6}$, outperforming
the trivial strategies described above. This was proved later to
be the optimal value \cite{gis97,bru98a,gis98}; in their original
paper, Bu\v{z}ek and Hillery had proved the optimality of their
transformation with respect to two different figures of merit.

\subsubsection{The transformation of the ancilla: "anti-clone"}
\label{ssanti}

Although everything was designed by paying attention to qubits A
and B, the partial state of the ancilla turns out to have a quite
interesting meaning too. We have \ba \rho_M &=&
\frac{2}{3}\ket{\psi^\perp}\bra{\psi^\perp}+\frac{1}{3}
\ket{\psi}\bra{\psi}= \frac{1}{2}\,\Big(\one -
\frac{1}{3}\hat{m}\cdot\vec{\si}\Big). \label{rhochb}\ea This
state is related to another operation which, like cloning, is
impossible to achieve perfectly, namely the NOT operation that
transforms $\ket{\psi}=\alpha\ket{0}+ \beta\ket{1}$ into
$\ket{\psi^{\perp}}=\alpha^*\ket{1}-\beta^*\ket{0}$. Because of
the need for complex conjugation of the coefficients, the perfect
NOT transformation is anti-unitary and cannot be
performed\footnote{Here is an intuitive version of this
impossibility result: any unitary operation on a qubit acts as a
rotation around an axis in the Bloch sphere, while the NOT is
achieved as the point symmetry of the Bloch sphere through its
center. Obviously, no rotation around an axis can implement a
point symmetry. A rotation of $\pi$ around the axis $z$ achieves
the NOT only for the states in the $(x,y)$ plane, while leaving
the eigenstates of $\si_z$ invariant.}. Just as for cloning, one
can choose to achieve the NOT on some states while leaving other
states unchanged; or one can find the operation that approximates
at best the NOT on all states, called the universal NOT (U-NOT).
This operation was anticipated in a remark by \textcite{bec99},
then fully described by \textcite{buz99}. The U-NOT gate gives
precisely $\rho_{NOT}=\rho_M$, and is thus implemented as a
by-product of cloning\footnote{Contrary to cloning however (see
the first trivial cloning strategy described above \ref{ssstriv}),
the optimal fidelity for the NOT {\em can} be reached also in a
measurement-based scenario \cite{buz99}.}. It has become usual to
say that, at the output of a QCM, the ancilla carries the optimal
{\em anti-clone} of the input state.

\subsection{Symmetric UQCM $N\rightarrow M$}
\label{sssymnm}

A symmetric, universal $N\rightarrow M$ QCM for qubits that
generalizes the B-H QCM was found by \textcite{gis97}. Its
fidelity is \ba F_{N\rightarrow
M}\,=\,\frac{MN+M+N}{M(N+2)}&\;&\;(d=2) \label{fqubits}\ea that
reproduces $F_{1\rightarrow 2}=\frac{5}{6}$ for $N=1$ and $M=2$.
They gave numerical evidence for its optimality. Later, an
analytical proof of optimality was given by \textcite{bru98b}, who
assumed that the output state belongs to the symmetric subspace of
$M$ qubits (this assumption is unjustified {\em a priori} but
turns out to be correct, see below). The result was further
generalized by Werner for systems of any dimension
\cite{wer98,key99}.

\subsubsection{Werner's construction}
\label{sswern}

We consider $d$-dimensional quantum systems described by the
Hilbert space ${\cal H}=\compl^d$. We introduce the notation
${\cal H}_+^n$ for the symmetric subspace of the $n$-fold tensor
product ${\cal H}^{\otimes n}$; the dimension of ${\cal H}_+^n$ is
$d[n]=\left(\begin{array}{c} d+n-1\\ n\end{array}\right)$. The
input state is $\rho_N=\sigma^{\otimes N}\,\in\,{\cal H}_+^N$,
where $\sigma=\ket{\psi}\bra{\psi}$ is a pure state. The QCM is
described by a trace preserving CP-map $T:{\cal H}_+^N\rightarrow
{\cal H}^{\otimes M}$. The remarkable fact is that one can
restrict to CP maps whose output is in the symmetric subspace
${\cal H}_+^M$. This is clearly true if one considers
"all-particle test criteria", such as minimizing the
trace-distance between $\rho_M$ and $\sigma^{\otimes M}$ or
maximizing the all-particle fidelity
$F_{all}=\mbox{Tr}\big[\sigma^{\otimes M}\,\rho_M\big]$, as
figures of merit \cite{wer98}; but if one wants to optimize the
single-copy fidelity, the restriction to the symmetric subspace is
not apparent at all, and required further work before being
demonstrated \cite{key99}.

In any case, the difficulty of the optimality proofs should not
hide the simplicity of the result: a single $T$ optimizes all the
figures of merit that have been considered, and this $T$ is in
some sense the most intuitive one. One simply takes the
non-symmetric trivial extension $\rho_N\rightarrow\rho_N\otimes
\one_{M-N}$, symmetrizes it and normalizes the result. Explicitly,
the optimal symmetric UQCM for $N\rightarrow M$ cloning reads \ba
T[\rho_N] &=& \frac{d[N]}{d[M]}\,S_M\, \big(\rho_N\otimes
\one_{M-N} \big)\,S_M \label{clonwerner}\ea where $S_M$ is the
projector from ${\cal H}^{\otimes M}$ to ${\cal H}_+^{M}$. The
constant \ba \frac{d[N]}{d[M]}&=&\left[\mbox{Tr} \Big(S_M\,
\big(\rho_N\otimes \one_{M-N}
\big)\,S_M\Big)\right]^{-1}\label{trace}\ea ensures that the map
$T$ is trace-preserving. The state of each clone is of the form
\ba \rho_1&=&
\eta(N,M)\ket{\psi}\bra{\psi}+[1-\eta(N,M)]\frac{\one}{d} \ea
where $\ket{\psi}$ is the input state and where the shrinking
factor is found to be \ba
\eta(N,M)&=&\frac{N}{M}\frac{M+d}{N+d}\,. \label{eta}\ea The
corresponding fidelity is \ba F_{N\rightarrow M}(d)&=&
\frac{N}{M}\,+\,\frac{(M-N)(N+1)}{M(N+d)}\,. \label{eq:fidwern}\ea
So this is the optimal fidelity for universal symmetric
$N\rightarrow M$ cloning of $d$-dimensional systems. For qubits
($d=2$) it indeed recovers the Gisin-Massar result
(\ref{fqubits}). For $N=1$ and $M=2$, $F=\frac{d+3}{2(d+1)}$. Note
that for a fixed amplification ratio $r=\frac{M}{N}$, the fidelity
goes as $F= 1-(d-1)\left(1-\frac{1}{r}\right)
\frac{1}{N}+O(N^{-2})$ for $N\rightarrow \infty$. Conversely, if
out of a finite number $N$ of originals one wants to obtain an
increasingly large number $M$ of clones, the fidelity of each
clone decreases as $F\simeq \frac{N+1}{N+d}$ in the limit
$M\rightarrow \infty$, in agreement with the results of state
estimation \cite{mas95} --- more in \ref{ss4meas} below.

\subsubsection{Calculation of the fidelity}

We have just summarized, without any proof, the main results for
the optimal universal symmetric $N\rightarrow M$ QCM with discrete
quantum systems. It is a good exercise to compute the single-copy
fidelity $F=\mbox{Tr}\left((\sigma\otimes\one...\otimes\one)
\,T[\sigma^{\otimes N}]\right)$ and recover (\ref{eq:fidwern}).
The first step is a symmetrization: denoting $\sigma^{(k)}$ the
operator that acts as $\sigma$ on the $k$-th system and as the
identity on the others, and replacing $T$ by its explicit form
(\ref{clonwerner}), we have \ban
F&=&\frac{d[N]}{d[M]}\,\frac{1}{M}\sum_{k=1}^M
\mbox{Tr}\left(\sigma^{(k)} \,S_M\, \big(\sigma^{\otimes N}\otimes
\one_{M-N} \big)\,S_M\right)\\ &=&
\frac{d[N]}{d[M]}\,\frac{1}{M}\sum_{k=1}^M \mbox{Tr}\left(S_M
\sigma^{(k)} \,\big(\sigma^{\otimes N}\otimes \one_{M-N}
\big)\,S_M\right)\ean where the second equality is obtained using
the linear and cyclic properties of the trace\footnote{Since the
trace is linear, we can bring the sum into it, then use
$\big(\sum_{k} \sigma^{(k)}\big)\,S_M=S_M\,\big(\sum_{k}
\sigma^{(k)}\big)$ and finally the cyclic properties of the
trace.}. Now, since $\si$ is a projector, $\sigma^{(k)}
\,\big(\sigma^{\otimes N}\otimes \one_{M-N} \big)$ is equal to
$\sigma^{\otimes N}\otimes \one_{M-N}$ for $1\leq k\leq N$; and is
equal to $\sigma^{\otimes N+1}\otimes \one_{M-N-1}$ for $N+1\leq
k\leq M$, where the additional $\sigma$ happens at different
positions. However, this is not important, since the expression is
sandwiched between the $S_M$ so it will be symmetrized anyway.
Using (\ref{trace}) and some algebra, one obtains
(\ref{eq:fidwern}).

\subsubsection{Trivial cloning revisited}
\label{ssstr2}

We can now have a different look at trivial cloning. The "trivial
amplification" strategy described in paragraph \ref{ssstriv} can
be easily generalized to the general case: one forwards the
original $N$ particles, adds $M-N$ particles prepared in the
maximally mixed state $\one/d$, and performs an {\em incoherent}
symmetrization (i.e., instead of projecting into the symmetric
subspace, one simply "shuffles" the particles). The fidelity is
then \ba F_{triv}(N\rightarrow M,d)&=& \frac{N}{M}
+\frac{M-N}{d\,M}\,.\label{ftriv}\ea As expected,
Eq.~(\ref{eq:fidwern}) shows that the Werner construction performs
better, but the difference vanishes in the limit $d\rightarrow
\infty$. We have thus learnt two new insights on optimal cloning:
(i) it is the quantum symmetrization that makes optimal cloning
non-trivial, and (ii) in the limit of large Hilbert space
dimension, trivial cloning performs almost optimally.

In summary, Werner's construction solves the problem of finding
the optimal universal symmetric QCM for any finite-dimensional
quantum system and for any number of input ($N$) and output
($M>N$) copies. We note that Werner did not provide the
implementation of the QCM $T$ as a unitary operation on the system
plus an ancilla (\ref{qcm}). This was provided by
\textcite{fan01}, generalizing previous partial results
\cite{buz98a,alb00}.  In the rest of this Section, we move to the
study of asymmetric (\ref{ssasym}) and state-dependent (i.e.,
non-universal) QCM.

\subsection{Asymmetric UQCM $1\rightarrow 1+1$}
\label{ssasym}

Asymmetric universal cloning refers to a situation where output
clones possibly have different fidelities. Here we focus on
$1\rightarrow 1+1$ universal cloning. The study of more general
cases has been undertaken recently \cite{ibl04,ibl05,fiu05},
motivated by the security analysis of practical quantum
cryptography \cite{aci04a,cur04}; we shall present some of these
ideas below, together with their possible experimental realization
(\ref{sssasymexp}).

In their comprehensive study of the $1\rightarrow 1+1$ cloning,
\textcite{niu98} had derived, in particular, the optimal
asymmetric UQCM $1\rightarrow 1+1$. The same result was found
independently by \textcite{cer98,cer00a} who used an algebraic
approach, and by \textcite{buz98b} who instead developed a quantum
circuit approach, improving over a previous construction for
symmetric cloning \cite{buz97}. Optimality is demonstrated by
proving that the fidelities of two clones, $F_A$ and $F_B$,
saturate the no-cloning inequality\footnote{This inequality
appears in all the meaningful papers with different notations. For
example, in \textcite{buz98a} it is Eq.~(11) since
$s_{0,1}=2F_{A,B}-1=1-2(1-F_{A,B})$; in \textcite{cer00a} it is
Eq.~(6), since $F_A=1-2x^2$ and $F_B=1-2x'^2$.} \ba
\sqrt{(1-F_A)(1-F_B)}&\geq & \demi-(1-F_A)-(1-F_B)\,.
\label{clonineq} \ea The same authors extended their constructions
beyond the qubit case to any $d$ \cite{cer00b,bra01}, although
optimality was only conjectured and was proved only recently
\cite{ibl04,ibl05,fiu05}.

We review both the Cerf's and the quantum circuit approaches,
giving the explicit formalism for qubits and explaining how this
generalizes to any dimension. We start with the quantum circuit
formalism, which is somehow more intuitive.

\subsubsection{Quantum circuit formalism}

\begin{center}
\begin{figure}
\includegraphics[width=7cm]{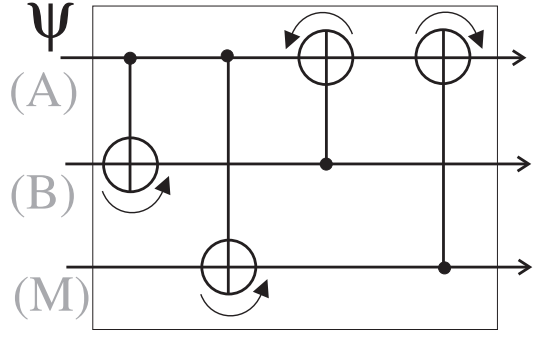}
\caption{The quantum circuit
used for universal $1\rightarrow 1+1$ cloning. For $d=2$, all the
gates are the standard CNOT. For $d>2$, the arrows play a role:
the arrow towards the right, resp. left, defines the
transformation $\ket{k}\ket{m}\rightarrow \ket{k}\ket{m+k}$, resp.
$\ket{k}\ket{m}\rightarrow \ket{k}\ket{m-k}$ --- as usual, sums
and differences in the kets are modulo $d$.} \label{figasym}
\end{figure}
\end{center}

The quantum circuit that is used for universal $1\rightarrow 1+1$
cloning in any dimension, which has been called a {\em quantum
information distributor}\footnote{This quantum circuit is
interesting beyond the interests of quantum cloning. Specifically,
\textcite{hil04} have identified in it a {\em universal
programmable quantum processor}. In short, the idea is to have a
circuit of logic gates coupling an input state with an ancilla,
such that any operation on the input state is obtained by a
convenient choice of the ancilla state (the "program"). No such
circuit exists if one requires it to work deterministically; the
present circuit does the job probabilistically (one knows when the
operation has succeeded).}, is drawn in Fig.~\ref{figasym}. It
uses a single $d$-dimensional system as ancilla. Let's focus on
qubits first. For states $\ket{\si}_A\ket{\omega}_B\ket{\xi}_M$ in
the computational basis, i.e. $\si,\omega,\xi\in\{0,1\}$, the
action of the circuit is \ba \ket{\si}_A\ket{\omega}_B\ket{\xi}_M
&\rightarrow & \ket{\si+\omega+\xi}_A\ket{\si+\omega}_B
\ket{\si+\xi}_M \ea where all the sums are modulo 2. It is now an
easy exercise to verify that \ba \ket{\psi}_A\ket{\Phi^+}_{BM}
&\rightarrow & \ket{\psi}_A\ket{\Phi^+}_{BM}\\
\ket{\psi}_A\ket{0}_B\ket{+}_{M} &\rightarrow &
\ket{\psi}_B\ket{\Phi^+}_{AM} \ea where
$\ket{\Phi^+}=\frac{1}{\sqrt{2}} \big(\ket{00}+\ket{11}\big)$ and
$\ket{+}=\frac{1}{\sqrt{2}} \big(\ket{0}+\ket{1}\big)$.
Figuratively, one can say that the state of B-M acts as the
"program" for the "processor" defined by the circuit; in
particular, $\ket{\Phi^+}_{BM}$  makes the processor act as the
identity on A; $\ket{0}_B\ket{+}_{M}$ makes the processor swap the
state $\ket{\psi}$ into mode B. Now, the optimal asymmetric QCM
follows quite intuitively: just take as an input state a coherent
superposition of "all the information in A" and "all the
information in B": \ba
\ket{\psi}_A\ket{\Psi_{in}}_{BM}&=&\ket{\psi}_A\,\big(a\ket{\Phi^+}_{BM}
+ b \ket{0}_B\ket{+}_{M}\big)\nonumber\\ &\rightarrow &
a\ket{\psi}_A\ket{\Phi^+}_{BM}\,+\,b
\ket{\psi}_B\ket{\Phi^+}_{AM}\,. \label{asymexpl}\ea The
parameters $a$ and $b$ are real; for the input state of B-M to be
normalized, they must satisfy $a^2+b^2+ab=1$. The partial states
for the two clones after the transformation read $\rho_{A,B}=
F_{A,B}\ket{\psi}\bra{\psi} +
(1-F_{A,B})\ket{\psi^{\perp}}\bra{\psi^{\perp}}$ where the
fidelities are \ba F_A\,=\,1-b^2/2 &,&F_B\,=\, 1-a^2/2\,.
\label{asymfid}\ea It is easy to verify that these fidelities
saturate the no-cloning inequality (\ref{clonineq}). As expected,
for $b=0$ (resp. $a=0$) we find all the information in A, resp. B.
The symmetric case corresponds to $a=b=\frac{1}{\sqrt{3}}$, in
which case we recover the Bu\v{z}ek-Hillery result
$F_A=F_B=\frac{5}{6}$.

The generalization to $d>2$ goes exactly along the same lines. The
state $\ket{\Phi^+}_{BM}$ is now the maximally entangled state of
two qudits $\frac{1}{\sqrt{d}}\sum_{k=0}^{d-1}\ket{k}_B\ket{k}_M$,
the state $\ket{+}_M$ is the superposition\footnote{Written
$\ket{p_0}$ in \textcite{bra01}, see Eq.~(2.1) in that reference.}
$\frac{1}{\sqrt{d}} \sum_{k=0}^{d-1}\ket{k}_M$. After the
transformation, the partial states of the two clones read
$\rho_{A}= (1-b^2)\ket{\psi}\bra{\psi} + b^2\one/d$ and $\rho_{B}=
(1-a^2)\ket{\psi}\bra{\psi} + a^2\one/d$, from which the
fidelities \ba F_A\,=\,1-\frac{d-1}{d}\,b^2 &,&F_B\,=\,
1-\frac{d-1}{d}\,a^2\,.\ea The normalization condition now reads
$a^2+b^2+\frac{2ab}{d}=1$; in particular, for the symmetric case
we recover Werner's result $F_A=F_B =\frac{d+3}{2(d+1)}$, see
Eq.~(\ref{eq:fidwern}).

\subsubsection{Cerf's formalism}

Cerf's formalism also uses a third $d$-dimensional system as
ancilla. For qubits, the transformation reads \ba
\ket{\psi}_A\ket{\Phi^+}_{BM}&\rightarrow & \ket{\Psi}_{ABM}
\,=\,V\,\ket{\psi}_A\ket{\Phi^+}_{BM} \label{cerftrans}\ea where
\ba V &= & \Big[v\,\one
\,+\,x\,\sum_{k=x,y,z}(\si_k\otimes\si_k\otimes\one)\Big]\,
\label{cerfimplicit}\ea where the real coefficients $v$ and $x$
must satisfy $v^2+3x^2=1$ to conserve the norm. Note that $V$, as
written here, is not unitary; however, (\ref{cerftrans}) defines a
unitary transformation. In other words, $V$ is the restriction of
a unitary operation when acting on input states of the form
$\ket{\psi}_A \ket{\Phi^+}_{BM}$. Here lies the appeal of Cerf's
formalism: the unitary that defines the QCM reduces to the very
compact and easily written transformation (\ref{cerfimplicit})
when acting on suitable input states. By inspection, one can
verify that the state $\ket{\Psi}$ in the r.h.s. of
(\ref{cerftrans}) is equal to (\ref{asymexpl}) with the
identification $a=v-x$, $b=2x$. In particular, the identity is
$v=1, x=0$, the swap is $v=x=\demi$, and the symmetric QCM is
$v=3x$, that is $x=\frac{1}{2\sqrt{3}}$.

The generalization to $d>2$ goes along the same lines
\cite{cer00b,cer02a}. The transformation, acting on the
$\ket{\psi}_A\ket{\Phi^+}_{BM}$ as defined above for qudits, reads
\ba V &= & \Big[v\,\one \,+\,x\,\sum_{(m,n)\in K}(U_{m,n}\otimes
U_{m,n}\otimes\one)\Big] \label{cerfimplicit2}\ea where
$K=\{(m,n)|0\leq m,n\leq d-1\}\setminus \{(0,0)\}$, from which the
normalization condition $v^2+(d^2-1)x^2=1$, and in which the
unitary operations $U_{m,n}$ that generalize the Pauli matrices
are defined as\footnote{In $\ket{k+m}$, the sum is modulo $d$.
Note also that, in the notation of \textcite{cer02a}, the
transformation (\ref{cerfimplicit2}) is written using
$U_{m,n}\otimes\one\otimes U_{m,-n}$ instead of $U_{m,n}\otimes
U_{m,n}\otimes\one$. This is indeed the same, since
$U_{m,-n}=U_{m,n}^*$ and it is well-known that
$U\otimes\one\ket{\Phi^+}=\one\otimes U^*\ket{\Phi^+}$ holds for
the maximally entangled state $\ket{\Phi^+}$.} \ba
U_{m,n}&=&\sum_{k=0}^{d-1}e^{2\pi i(kn/d)}\ket{k+m}\bra{k}\,. \ea
The link with the parameters $a$ and $b$ of the quantum circuit
formalism is provided here by\footnote{To derive this, replace
$F=F_A$ given in (\ref{asymfid}) into Eq.~(16) of
\textcite{cer02a}.} $a=v-x$, $b=dx$.

\subsection{State-dependent cloning}
\label{ssstate}

\subsubsection{Cloning of two states of qubits}

The first study of state-dependent cloning was based on a
different idea, simply, to clone at best two arbitrary pure states
of a qubit \cite{bru98a}. This is a hard problem because of the
lack of symmetry, and was not pursued further. One wants to
perform the optical symmetric cloning of two states of qubits
$\ket{\psi_0}$ and $\ket{\psi_1}$, related by
$|\braket{\psi_0}{\psi_1}|=s$. The resulting fidelity for this
task is given by a quite complicated formula: \ba
F&=&\demi+\frac{\sqrt{2}}{32s}(1+s)
\left(3-3s+\sqrt{1-2s+9s^2}\right)\nonumber \\&\times&
\sqrt{-1+2s+3s^2+(1-s)\sqrt{1-2s+9s^2}}\,. \ea For $s=0$ and
$s=1$, one finds $F=1$ as it should, because the two states belong
to the same orthogonal basis. The minimum is $F\approx 0.987$,
much better than the value obtained with the symmetric
phase-covariant cloner (see below). Oddly enough, this minimum is
achieved for $s=\demi$, while one would have expected it to occur
for states belonging to mutually unbiased bases
($s=\frac{1}{\sqrt{2}}$).

\subsubsection{Phase-covariant $1\rightarrow 2$ for qubits: generalities}

The best-known example of state-dependent QCM are the so-called
{\em phase-covariant QCM}. For qubits, these are defined as the
QCM that copy at best states of the form \ba
\ket{\psi(\varphi)}&=&\frac{1}{\sqrt{2}}\left(\ket{0}+
e^{i\varphi}\ket{1}\right)\,. \ea These are the states whose Bloch
vector lies in the equator ($x-y$) of the Bloch sphere; the name
"phase-covariant", used for the first time by \textcite{bru00b},
comes from the fact that the fidelity of cloning will be
independent of $\varphi$. Here we restrict our attention to
$1\rightarrow 2$ asymmetric phase-covariant cloning for qubits.

The phase-covariant QCM has a remarkable application in quantum
cryptography, since it is used in the optimal incoherent strategy
for eavesdropping on the BB84 protocol, see \ref{sssopt} below.
Note that the eavesdropper on BB84 wants to gather information
only on four states, defined by
$\varphi=0,\frac{\pi}{2},\pi,\frac{3\pi}{2}$: the eigenstates of
$\si_x$ and $\si_y$, that is, two maximally conjugated bases. But
the two problems (cloning all the equator, or cloning just two
maximally conjugated bases on it) yield the same solution. In
fact, consider a machine, a CP map $T$, that clones optimally the
four states of BB84, in the sense that when acting on $\ket{\pm
x}$ and $\ket{\pm y}$, it gives two approximate clones of the
form\footnote{Notice that it is assumed here that the cloning
process only shrinks the Bloch vector of the initial input state.}
\begin{eqnarray}
  T[\ket{\pm x}\bra{\pm x}] &=& \eta\ket{\pm x}\bra{\pm x}+(1-\eta)\frac{\one}{2}
  \nonumber\\
  T[\ket{\pm y}\bra{\pm y}] &=&
  \eta\ket{\pm y}\bra{\pm y}+(1-\eta)\frac{\one}{2} .
\end{eqnarray}
Any state in the equator of the Bloch sphere can be written as
\begin{equation}
    \ket{\psi(\varphi)}\bra{\psi(\varphi)}=\frac{1}{2}
    \left(\one+\cos\varphi\,\sigma_x+\sin\varphi\,\sigma_y
    \right)
\end{equation}
Now, using the linearity of $T$ one can see that
$T[\sigma_x]=\eta\sigma_x$ and the same holds for $\sigma_y$.
Since $T(\one)=\one$, one has \ba
T[\ket{\psi(\varphi)}\bra{\psi(\varphi)}]&=&\eta
\ket{\psi(\varphi)}\bra{\psi(\varphi)}+(1-\eta)\frac{\one}{2} \ea
for all $\varphi$. This shows that the optimal cloning of the four
states employed in the BB84 protocol is equivalent to optimally
cloning the whole equator of the Bloch sphere. A similar argument
applies if the $z$ basis is also included: to clone all mutually
unbiased bases in the Bloch sphere, i.e. the states $\ket{\pm x}$,
$\ket{\pm y}$ and $\ket{\pm z}$, is equivalent to universal
cloning.

\subsubsection{Phase-covariant $1\rightarrow 2$ for qubits: explicit transformation}

The task of copying at best the equator of the Bloch sphere, even
in the asymmetric case, can be accomplished {\em without ancilla}
\cite{niu99}; this is definitely impossible for universal cloning
\cite{dur03b}. The QCM is then just part of a two-qubit unitary
transformation that reads \ba
\begin{array}{ccl} \ket{0}\ket{0} &\rightarrow &
\ket{0}\ket{0}\\
\ket{1}\ket{0} &\rightarrow & \cos\eta\,\ket{1}\ket{0}\,+\,
\sin\eta\,\ket{0}\ket{1}\end{array} \label{uning}\ea with $\eta\in
\left[0,\frac{\pi}{2}\right]$ and we have chosen
$\ket{R}=\ket{0}$. Then $\ket{\psi(\varphi)}\ket{0}\rightarrow
\frac{1}{\sqrt{2}}\left(\ket{0}\ket{0} +\cos\eta
e^{i\varphi}\ket{1}\ket{0}+ \sin\eta
e^{i\varphi}\ket{0}\ket{1}\right)$; the partial states $\rho_A$
and $\rho_B$ are readily computed, and one finds the fidelities
\ba F_A^{x-y}\,=\,\demi\big(1+\cos\eta\big)\,&,&\,
F_B^{x-y}\,=\,\demi\big(1+\sin\eta\big)\,. \label{fidxy}\ea As
desired, these fidelities are independent of $\varphi$. It is
easily verified numerically that this QCM is better than the
universal one for the equatorial states: one simply fixes
$F_A=F_A^{x-y}$ and verifies that $F_B^{x-y}\geq F_B$, where $F_B$
is given in (\ref{asymfid}). In particular, for the symmetric case
$\eta=\frac{\pi}{4}$, one has $F_{A,B}^{x-y}=
\demi\big(1+\frac{1}{\sqrt{2}}\big)\simeq 0.8535
\,>\,\frac{5}{6}$.

Niu and Griffiths introduced the two-qubit QCM in the context of
eavesdropping in cryptography. It was later realized that a
version with ancilla of the phase-covariant QCM
\cite{gri97,bru00b}, while equivalent in terms of fidelity of the
clones on the equator, is generally more suited for the task of
eavesdropping \cite{dur03b,aci04a,aci04b}. This machine can be
constructed by symmetrizing (\ref{uning}) with the help of an
ancilla qubit as follows: \ba
\begin{array}{ccl} \ket{0}\ket{0}\ket{0} &\rightarrow &
\ket{0}\ket{0}\ket{0}\\
\ket{1}\ket{0}\ket{0} &\rightarrow &
\big(\cos\eta\,\ket{1}\ket{0}\,+\, \sin\eta\,\ket{0}\ket{1}\big)\,\ket{0} \\
\ket{0}\ket{1}\ket{1} &\rightarrow &
\big(\cos\eta\,\ket{0}\ket{1}\,+\, \sin\eta\,\ket{1}\ket{0}\big) \, \ket{1}\\
\ket{1}\ket{1}\ket{1} &\rightarrow &
\ket{1}\ket{1}\ket{1}\end{array}\label{unicerf}\ea and letting
this unitary act on the input state
$\ket{\psi}_A\ket{\Phi^+}_{BM}$. This reminds Cerf's formalism,
and indeed the unitary (\ref{unicerf}) acts on
$\ket{\psi}_A\ket{\Phi^+}_{BM}$ as the operator \cite{cer00b}\ba
V&=&F\,\one_{ABM} \,+\,
(1-F)\si_{z}\otimes\si_{z}\otimes\one\nonumber\\&&+\,
\sqrt{F(1-F)}\,\big(\si_{x}\otimes\si_{x}+
\si_{y}\otimes\si_{y}\big)\otimes\one \label{cerfpc}\ea where
$F=F_{A}^{x-y}$. Notice again how Cerf's formalism appeals to
intuition: it is manifest in (\ref{cerfpc}) that the $x-y$ plane
is treated differently from the $z$ direction. For a practical
illustration of the use of the phase-covariant QCM for
eavesdropping in cryptography, we refer the reader to paragraph
\ref{sssopt} below.

\subsubsection{Other state-dependent QCM}

Most of the state-dependent QCM that have been studied are
generalizations of the phase-covariant one, often called
phase-covariant as well. The idea is to clone at best some
maximally conjugated bases. Specifically, the following
state-dependent cloners have been studied:

\begin{itemize}

\item Asymmetric $1\rightarrow 2$ phase-covariant QCM, that clones
at best two maximally conjugated bases in any dimension
\cite{cer02a,fan03}. For $d=3$ \cite{cer02b} and $d=4$
\cite{dur03a}, asymmetric $1\rightarrow 2$ QCM have been provided
that clone three or four maximally conjugated bases. For any $d$,
the symmetric QCM that are optimal for cloning "real quantum
states" --- that is, a basis and all the states obtained from it
using $SO(d)$ --- have been found; optimality has been
demonstrated using the no-signaling condition \cite{nav03}.

\item Symmetric $N\rightarrow M$ phase-covariant QCM for arbitrary
dimension \cite{bus04}, generalizing previous results
\cite{dar03}. In particular, machines have been found that work
without ancilla ("economical QCM") thus generalizing the
Niu-Griffiths construction  given above (\ref{uning}) --- which,
however, provides also the asymmetric case.

\item Not related to phase-covariant cloning: \textcite{fiu02}
have studied the cloning of two orthogonal qubits. It is known
that, for the task of estimating a direction $\hat{n}$, the two
qubit state $\ket{\hat{n},-\hat{n}}$ gives a better estimate than
the state $\ket{\hat{n},\hat{n}}$ \cite{gis99}. For cloning, the
task is to produce $M$ clones of $\ket{\hat{n}}$ starting from
either of those two-qubit states. For $M\leq 6$, better copies are
obtained when starting from $\ket{\hat{n},-\hat{n}}$.

\item Finally, another issue that has been discussed is the
optimal cloning of entangled states \cite{lam04}.

\end{itemize}

\subsection{Quantum cloning and state estimation}
\label{ss4meas}

One could anticipate that there might exist a strong relation
between cloning the state of a quantum system and acquiring
knowledge about this state. After all, there is a strong analogy
between the two processes. In both cases, the (quantum)
information contained in the input is transferred into some
"larger" system: the output clones in the case of cloning, and the
measuring device in the case of state estimation. In this section,
we shall see that there is more than a mere analogy. In fact, as
first appreciated by \textcite{gis97} and further elaborated by
\textcite{bru98b}: (i) There is an equivalence between optimal
universal $N \to \infty$ quantum cloning machines of pure states
and optimal state estimation devices taking as input $N$ replicas
of an unknown pure state. (ii) Bounds on optimal cloning can be
derived from this equivalence. We are going to present these
results. To simplify the presentation, we shall only consider
universal cloning of qubits, but the subsequent analysis can be
generalized, without difficulty, to qudits \cite{key02}.

The equivalence between optimal universal symmetric cloning and
state estimation can be established using the notion of {\em
shrinking factor}, already introduced for cloners. Indeed, we
stressed in paragraph \ref{sssymnm} that the quality of the
optimal $N \to M$ UQCM is fully characterized by its shrinking
factor $\eta(N,M)$: if the input state to clone reads
$\rho_{\textrm{in}}=\ket{\psi}\bra{\psi}$, then the individual
state of each output clone reads $\eta(N,M)
\ket{\psi}\bra{\psi}+(1-\eta(N,M)) \frac{\openone}{2}$. A similar
structure arises in the case of state estimation. Given $N$ copies
of an unknown qubit state $\ket{\psi}$, there exists an optimal
POVM (Positive-Operator-Valued Measure) \ba\label{eq:povmmas95}
P_{\mu} \geq 0\;&,&\; \sum_{\mu} P_{\mu}= S_N \ea
which\footnote{Note that the elements of the POVM sum up to the
projector $S_N$ onto the symmetric subspace ${\cal H}^N_+$, and
not to the identity. In fact, one can "complete" the POVM with the
operator $\one-S_N$, but the corresponding outcome will never be
observed because the input state belongs to ${\cal H}^N_+$.}
yields the best possible estimate of $\psi$ taking the fidelity as
figure of merit \cite{mas95}. To each measurement outcome $\mu$, a
guess $\ket{\psi_{\mu}}$ of the input state is associated. During
\emph{one} instance of the state estimation experiment, the
outcome $\mu$ can appear with probability $\tr P_{\mu}
\ket{\psi}\bra{\psi}=p_{\mu}(\psi)$. Thus, \emph{on average}, the
POVM (\ref{eq:povmmas95}) yields the estimate
$\rho_{\textrm{est}}(\psi)=\sum_{\mu} p_{\mu}(\psi)
\ket{\psi_{\mu}}\bra{\psi_{\mu}}$. It turns out \cite{mas95} that
this average estimate can be written as \beq\label{eq:stateest}
\eta_*(N) \ket{\psi}\bra{\psi}+(1-\eta_*(N)) \frac{\openone}{2},
\eeq and the average fidelity of the state estimation is thus
given by $(1+\eta_*(N))/2$. In turn, the performance of the POVM
that describes the best state estimation can also be characterized
by a shrinking factor $\eta_*(N)$.

We can now state precisely what we mean when stating that there
exists an equivalence between optimal $N \to \infty$ quantum
cloning machine and an optimal state estimation device. We have
\beq\label{eq:shrinkid} \eta_*(N)=\eta(N,\infty). \eeq This
relation tells us that using $N$ qubits identically prepared in
the state $\ket{\psi}$ to estimate $\psi$ or to prepare an
infinite number of clones of $\ket{\psi}$ (and then infer an
estimate of $\psi$) are essentially equivalent procedures: the
amount of information one can extract about the input preparation
is the same in both cases.

To prove Eq.~(\ref{eq:shrinkid}), we shall show that both
$\eta_*(N) \leq \eta(N,\infty)$ and $\eta_*(N) \geq
\eta(N,\infty)$ hold. The first of these inequalities is almost
obvious. Consider a $N \to M$ cloning procedure in which we first
perform state estimation on the $N$ input originals, and then
prepare $M$ output clones according to the (classical) outcome we
get.  If the input state is $\ket{\psi}^{\otimes N}$, then, on
average, the state of each clone will be of the form
(\ref{eq:stateest}), and thus characterized by a shrinking factor
$\eta_*(N)$.  By definition, such a cloning procedure cannot be
better than using an optimal $N \to M$ quantum cloning machine.
Thus $\eta_*(N) \leq \eta(N,M)$ for all $M$, and in particular
$\eta_*(N) \leq \eta(N,\infty)$.

To prove the second inequality, $\eta_*(N) \geq \eta(N,\infty)$,
we shall conversely consider a situation in which we want to
achieve state estimation from $N$ input originals with an
intermediate cloning step. Let us remark that the output of an
optimal $N \to M$ UQCM belongs to the symmetric subspace
$\hilb^N_+$. Therefore \cite{bru98b}, for any input state
$\ket{\psi}^{\otimes N}$, the output state can be written as a
pseudo-mixture \beq\label{eq:pseudomix} \sum_i \alpha_i(\psi)
\ket{\psi_i}\bra{\psi_i}^{\otimes M}, \eeq that is $\sum_i
\alpha_i(\psi)=1$ but the coefficients $\alpha_i(\psi)$ may be
negative. Also, from $\ket{\psi_i}\bra{\psi_i}^{\otimes M}$, our
optimal state estimation device yields (on average) the estimate $
\eta_*(M) \ket{\psi_i}\bra{\psi_i}+(1-\eta_*(M))
\frac{\openone}{2}$. Thus, by linearity, our estimation procedure
yields the estimate \bed \rho_{\textrm{est}}=\sum_i \alpha_i(\psi)
\left(\eta_*(M) \ket{\psi_i}\bra{\psi_i}+(1-\eta_*(M))
\frac{\openone}{2}\right). \eed Clearly, $\sum_i \alpha_i(\psi)
\ket{\psi_i}\bra{\psi_i}=\eta(N,M)
\ket{\psi}\bra{\psi}+(1-\eta(N,M)) \frac{\openone}{2}$. By
definition, this state estimation scheme cannot outperform an
optimal state estimation on the $N$ input originals. Thus
$\eta(N,M) \eta_*(M) \leq \eta_*(N)$. From the fact that in the
limit of large $M$ states estimation can be accomplished
perfectly, $\lim_{M \to \infty} \eta_*(M)=1$ \cite{mas95}, we
deduce that $\eta(N,\infty) \leq \eta_*(N)$. This concludes the
proof of Eq.~(\ref{eq:shrinkid}).

We are now in a position to further connect quantum cloning and
state estimation. Starting from (\ref{eq:shrinkid}), we can show
that a limit on the quality of $N \to M$ cloning can be derived
from state estimation, modulo one assumption: the output state of
an $N \to M$ cloning machine should be supported by the symmetric
subspace $\mathcal{H}_M^+$. To establish such a limit, our first
task is to prove that the shrinking factors of two cascaded
cloners multiply. Let us construct an $N \to L$ cloning machine by
concatenating an $N \to M$ machine with an $M \to L$ machine, and
let such a cloning machine act on some input state $\psi^{\otimes
N}$. Since the output state of the first cloner is assumed to be
supported by $\mathcal{H}_M^+$, it admits the decomposition
(\ref{eq:pseudomix}). Processing this output state into the second
cloning machine yields \beq \sum_j \beta_j(\psi_i)
\ket{\psi_j}\bra{\psi_j}^{\otimes L}, \eeq where $\sum_j
\beta_j(\psi_i) \ket{\psi_j}\bra{\psi_j}=\eta(M,L)
\ket{\psi_i}\bra{\psi_i}+(1-\eta(M,L)) \frac{\openone}{2}$. Thus
the individual state of each clone at the output of the second
cloner reads $\eta(N,M) \eta(M,L)\ket{\psi}\bra{\psi}+(1-\eta(N,M)
\eta(M,L)) \frac{\openone}{2}$. Of course, this cloning in stage
cannot be better than directly using an optimal $N \to M$ cloner.
Thus \bed \eta(N,M) \eta(M,L) \leq \eta(N,L). \eed In particular,
$\eta(N,M) \eta(M,\infty) \leq \eta(N,\infty)$. Using
Eq.~(\ref{eq:shrinkid}), we deduce the important relation
 \beq\label{eq:shrinkineq}
 \eta(N,M) \,\leq\, \frac{\eta_*(N)}{\eta_*(M)}\,.
 \eeq
From $\eta_*(N)=N/(N+2)$ \cite{mas95}, we find \be \eta(N,M)\,
\leq \,\frac{N}{M} \frac{M+2}{N+2}\,. \ee Comparing with
(\ref{eta}), we see that, perhaps not so surprisingly, this last
inequality is saturated by optimal UQCM.

The foregoing analysis establishes a precise connection between
optimal cloning and optimal state estimation, valid when one
considers all possible pure states of qubits ---- in fact, it
extends to all pure states of qudits for any $d$ --- and looks
like a miracle. One could argue that the main reason why this
connection appears is that the output state of an optimal $N \to
M$ cloning machine \emph{turns out} to be supported by the
symmetric subspace $\mathcal{H}_M^+$, the crucial ingredient in
deriving (\ref{eq:shrinkid}) and (\ref{eq:shrinkineq}). But this
latter fact, although established on a firm mathematical ground
\cite{key99} is still lacking a physical interpretation. A recent
result has come to strengthen this connection: it has been proved
\cite{ibl04,ibl05,fiu05} that the optimal asymmetric $1\rightarrow
1+N$ UQCM, in the limit $N\rightarrow \infty$, achieves the
optimal "disturbance vs. gain" trade-off for the measurement of
one qubit \cite{ban01}.

One might wonder if the connection between state estimation and
cloning holds in general. To our knowledge, the question is still
open. It certainly deserves further investigation, for answering
it would allow to understand whether the neat relation between
cloning and state estimation is a fundamental feature of quantum
theory or a mere peculiarity of the set of all pure states of
qudits.

\section{Cloning of continuous variables}
\label{sec3}

This Section reviews the issue of approximate cloning for
continuous variable systems (or quantum oscillators). Our analysis
will be focussed on $N \to M$ Gaussian machines, cloning equally
well all coherent states
\cite{cer00c,cer00d,lin00,bra01sof,fiu01}. The optimality of such
machines will be investigated. Upper bounds on the minimal amount
of noise the clones should feature will be derived for qubits
(\ref{sect:opticlongauss}) via a connection with quantum
estimation theory, using techniques similar to those we have
presented in paragraph \ref{ss4meas}. Then we shall present
transformations achieving these bounds (\ref{sec:cvimpl}).
Finally, we shall briefly discuss possible variants of our
analysis (\ref{sec:cvothers}).

\subsection{Optimal cloning of Gaussian states}\label{sect:opticlongauss}

\subsubsection{Definitions and results}

The Hilbert space associated with a quantum oscillator is
$\mathcal{H} \equiv L^2(\mathbf{R})$, and is infinite-dimensional.
Let us first consider what we can get from asking for universality
in such a Hilbert space. Considering the limit for $d \to \infty$
of Eq.~(\ref{eq:fidwern}), we see that \beq\label{eq:funidinf}
\lim_{d \to \infty} F_{N\rightarrow M}(d)=\frac{N}{M}, \eeq where
$N$ is the number of input replicas, and $M > N$ the number of
clones. Moreover, this limit can also be reached by trivial
cloning, see \ref{ssstr2}. Can we then do better than
Eq.~(\ref{eq:funidinf}), by dropping the requirement of
universality or taking a different perspective? After all, in some
circumstances such as quantum cryptography, it is natural to
consider cloners which are optimal only for a subset of states
$\mathcal{S} \subset \mathcal{H}$. Also, the fidelity is not
always the most interesting figure of merit to consider.

Here, we shall concentrate on the situation in which we only want
to clone the set of {\em coherent states}, denoted $\mathcal{S}$.
Let $\hat{x}$ and $\hat{p}$ denote two mutually conjugated
quadratures of a harmonic oscillator, $[\hat{x},\hat{p}]=i$
($\hbar=1$). The set of coherent states is the set of states that
satisfy \beq \Delta \hat{x}^2=\moy{\hat{x}^2}-\moy{\hat{x}}^2=
\Delta\hat{p}^2=\moy{\hat{p}^2}-\moy{\hat{p}}^2=1/2, \eeq and can
be parametrized as \beq \mathcal{S}=\left\{\ket{\alpha}:
\alpha=\frac{1}{\sqrt{2}}(x+ip), x,p \in \mathbf{R}\right\}, \eeq
where $\bra{\alpha}\hat{x}\ket{\alpha}=x$ and
$\bra{\alpha}\hat{p}\ket{\alpha}=p$. We shall consider $N \to M$
symmetric Gaussian cloners (SGC). These cloners are linear,
trace-preserving, completely positive maps $\mathcal{C}$
outputting $M$ clones from $N \leq M$ identical replicas of an
unknown coherent state $\ket{\alpha}$. To simplify the analysis,
we require that the joint state of the $M$ clones
$\mathcal{C}(\ket{\alpha}\bra{\alpha}^{\otimes N})$ be supported
on the symmetric subspace of $\mathcal{H}^{\otimes M}$ and be such
that the partial trace over all output clones but (any) one is the
bi-variate Gaussian mixture: \beqa \label{eq:reducrho}
&&\rho_1(\alpha) \,=\, {\mathrm Tr}_{M-1} \mathcal{C}(\ket{\alpha}\bra{\alpha}^{\otimes N}) \nonumber \\
&&\,=\,\frac{1}{\pi \sigma^2_{N,M}} \int d^2\beta
e^{-|\beta|^2\sigma^2_{N,M}}
D(\beta)\ket{\alpha}\bra{\alpha}D^{\dagger}(\beta) \eeqa where the
integral is performed over all values of $\beta=(x+ip)/\sqrt{2}$
in the complex plane ($\hbar=1$), and the operator
$D(\beta)=\exp(\beta a^\dagger -\beta^*  a)$ achieves a
displacement of $x$ in position and $p$ in momentum, with
$\hat{a}=\frac{1}{\sqrt{2}}(\hat{x}+i\hat{p})$ and
$\hat{a}^\dagger=\frac{1}{\sqrt{2}}(\hat{x}-i\hat{p})$ denoting
the annihilation and creation operators, respectively. Thus, the
copies yielded by a SGC are affected by an equal Gaussian noise
$\sigma_x^2=\sigma_p^2=\sigma_{N,M}^2$ on the conjugate variables
$x$ and $p$. The fidelity of the optimal $N \to M$ SGC when a
coherent state $\ket{\alpha}$ is copied can be computed using
Eq.~(\ref{eq:reducrho}) and the identity $|\langle
\alpha|\alpha'\rangle|^2=\exp(-|\alpha-\alpha'|^2)$. One finds
\beq\label{eq:bestfid} f_{N,M} =\bra{\alpha}\rho_1\ket{\alpha} =
\frac{1}{1+\overline{\sigma}^2_{N,M}}\,. \eeq We shall prove in
the following paragraphs that a lower bound on the noise variance
$\sigma_{N,M}^2$ is given by \beq\label{eq:mainresult}
\overline{\sigma}^2_{N,M} = \frac{1}{N}-\frac{1}{M}, \eeq implying
in turn that the optimal cloning fidelity for Gaussian cloning of
coherent states is bounded by \beq  \label{eq:fidelity}
f_{N,M}={MN\over MN+M-N}\,. \eeq Thus, all coherent states are
copied with the same fidelity --- recall that this property does
not extend to all states of ${\mathcal H}$. One can also check
that Eqs (\ref{eq:mainresult}) and (\ref{eq:fidelity}) fulfill the
natural requirement that the cloning fidelity increases with the
number of input replicas. At the limit $N \to \infty$, we have
$f_{N,M} \to 1$ for all $M$, that is, classical copying is
allowed. Finally, for $M \to \infty$, that is, for an optimal
measurement, we get $f_{N,M} \to N/(N+1)$.

It is worth noting that the optimal cloning of {\em squeezed
states} requires a variant of these SGCs. For instance, the best
symmetric cloner for the family of quadrature squeezed states with
squeezing parameter $r$ must have the form of
Eq.~(\ref{eq:reducrho}), but using the definition
$\beta=(\frac{x}{\kappa}+i\kappa p)/\sqrt{2}$ with
$\kappa=\exp(r)$. These cloners naturally generalize the SGCs and
give the same cloning fidelity, Eq.~(\ref{eq:fidelity}), for those
squeezed states.

\subsubsection{Proof of the bounds for $1\rightarrow 2$ cloning}

Let us first prove (\ref{eq:mainresult}) in the simplest case
$(N,M)=(1,2)$. This case is interesting to single out because it
demonstrates the link between quantum cloning and the problem of
simultaneously measuring a pair of conjugate observables on a
single quantum system. Our starting point is thus the relation
derived by \textcite{art65}, which constrains any attempt to
measure $\hat{x}$ and $\hat{p}$ simultaneously on a quantum
system: \beq \label{eq:ak} \sigma^2_x(1) \; \sigma^2_p(1) \geq 1,
\eeq where $\sigma^2_ x(1)$ and $\sigma^2_p(1) $ denote the
variance of the \emph{measured} values of $\hat{x}$ and $\hat{p}$,
respectively, when simultaneously measuring $\hat{x}$ and
$\hat{p}$ on some quantum state $\rho$.

It is crucial to clearly distinguish between the Arthurs and Kelly
relation (\ref{eq:ak}), and the Heisenberg uncertainty
relation\footnote{In this paper, we adopt the usual notation
$\Delta A$ for the intrinsic variance of the observable $A$, and
use $\delta A$ for other variances in Eq.~(\ref{delta}).
Obviously, $\delta A\geq \Delta A$; this is what motivates the use
of the opposite convention in the papers on cloning that we are
reviewing here.}: \beq\label{eq:heis} \Delta\hat{x}^2
\Delta\hat{p}^2 \geq 1/4, \eeq where $\Delta\hat{x}^2$ (resp.
$\Delta\hat{p}^2$) are \emph{intrinsic} variance of the
observables $\hat{x}$ (resp. $\hat{p}$) for any quantum state
$\rho$. The Heisenberg relation is valid independently from any
measurement performed on the state $\rho$; in particular, it holds
even if we have a perfect knowledge of the state $\rho$. In
contrast, the trade-off between the information about $\hat{x}$
and the information about $\hat{p}$, that one can acquire during a
single measurement on the state $\rho$, is quantified by the
Arthurs-Kelly relation (\ref{eq:ak}). In particular, the best
possible simultaneous measurement of $\hat{x}$ and $\hat{p}$ with
a same precision satisfies $\sigma^2_x(1)=\sigma^2_p(1)=1$.
Compared with the intrinsic noise of a coherent state
$\Delta\hat{x}^2=\Delta\hat{p}^2=1/2$, we see that the joint
measurement of $x$ and $p$ effects an additional noise of minimum
variance 1/2.

Now, let a coherent state $\ket{\alpha}$ be processed by a $1 \to
2$ SGC, and let $\hat{x}$ be measured at one output of the cloner
while $\hat{p}$ is measured at the other output. This is a way of
simultaneously measuring $x$ and $p$, and as such it must obey the
Arthurs-Kelly relation (\ref{eq:ak}). Consequently, the intrinsic
variances of the observable $\hat{x}$ and $\hat{p}$ in the state
$\rho_1(\alpha)$, denoted respectively as $\delta \hat{x}$ and
$\delta \hat{p}$, must fulfill \beq \delta \hat{x}^2 \; \delta
\hat{p}^2 \geq 1\,. \label{delta}\eeq Using
Eq.~(\ref{eq:reducrho}), we get \beq
(\Delta\hat{x}^2+\sigma_{1,2}^2) (\Delta\hat{p}^2+\sigma_{1,2}^2)
\geq 1. \eeq Now using Eq.~(\ref{eq:heis}), we conclude that the
noise variance is constrained by \beq \sigma^2_{1,2} \geq
\overline{\sigma}^2_{1,2} = 1/2, \eeq thus verifying
Eq.~(\ref{eq:mainresult}) in the case $(N,M)=(1,2)$.

A similar argument can be used to characterize the output copies
of an asymmetric quantum cloning machine, in which the qualities
of the clones are not identical and in which one might desire that
the added noise due to cloning is different for both quadratures.
Using, Eq.~(\ref{eq:ak}), one easily shows that the following
relations hold: \beqa\label{eq:clonuncert}
\sigma^2_{x,1} \sigma^2_{p,2} \geq 1/4, \\
\sigma^2_{p,1} \sigma^2_{x,2} \geq 1/4, \eeqa where
$\sigma^2_{x,1}$ (resp. $\sigma^2_{p,1}$) refers to the added $x$
quadrature (resp. $p$ quadrature) added noise for the first clone,
and where $\sigma^2_{x,2}$ and $\sigma^2_{p,2}$ are defined
likewise. These cloning uncertainty relations are useful when
assessing the security of some continuous variables quantum
cryptographic schemes \cite{cer01}.

\subsubsection{Proof of the bounds for $N\rightarrow M$ cloning}

Let us now prove Eq.~(\ref{eq:mainresult}) in the general case.
Our proof is connected to quantum state estimation theory
similarly to what was done for quantum bits in paragraph
\ref{ss4meas}. The key idea is that cloning should not be a way of
circumventing the noise limitation encountered in any measuring
process. More specifically, our bound relies, as in the discrete
case, on the fact that cascading an $N \to M$  cloner with an $M
\to L$ cloner results in a $N\to L$ cloner which cannot be better
that the {\em optimal} $N \to L$ cloner. We make use of the
property that cascading two SGCs results in a single SGC whose
variance is simply the sum of the variances of the two component
SGCs \cite{cer00c}. Hence, the variance
$\overline{\sigma}^2_{N,L}$ of the {\em optimal} $N \to L$ SGC
must satisfy \beq \overline{\sigma}^2_{N,L} \leq
\sigma^2_{N,M}+\sigma^2_{M,L}. \eeq In particular, if the $M\to L$
cloner is itself optimal and $L\to\infty$, \beq\label{eq:addvar}
\overline{\sigma}^2_{N,\infty} \leq \sigma^2_{N,M}+
\overline{\sigma}^2_{M,\infty} \eeq As for the discrete case, in
the limit $M \to \infty$, estimators and quantum cloning machines
tend to become essentially identical devices. Thus
Eq.~(\ref{eq:addvar}) means that cloning the $N$ replicas of a
system before measuring the $M$ resulting clones does not provide
a mean to enhance the accuracy of a direct measurement of the $N$
replicas.

Let us now estimate $\overline{\sigma}^2_{N,\infty}$, that is, the
variance of an optimal joint measurement of $\hat{x}$ and
$\hat{p}$ on $N$ replicas of a system. From quantum estimation
theory \cite{hol82}, we know that the variance of the measured
values of $\hat{x}$ and $\hat{p}$ on a single system, respectively
$\sigma^2_x(1)$ and $\sigma^2_p(1)$, are constrained by
\beq\label{eq:holevobound} g_x \sigma^2_x(1)+g_p \sigma^2_p(1)
\geq g_x \Delta \hat{x}^2+ g_p \Delta \hat{p}^2 + \sqrt{g_x g_p}
\eeq for all values of the constants $g_x, g_p >0$. Note that, for
each value of $g_x$ and $g_p$, a specific POVM based on a
resolution of identity in terms of squeezed states, whose
squeezing $\Delta$ is a function of $g_x$ and $g_p$, achieves this
bound \cite{hol82}. Squeezed states satisfy
$\Delta\hat{x}^2=\kappa^2/2$ and $\Delta\hat{p}^2=1/2\kappa^2$.
Moreover, when a measurement is performed on $N$ independent and
identical systems, the r.~h.~s. of (\ref{eq:holevobound}) is
reduced by a factor $N^{-1}$, as in  classical statistics
\cite{hel76}. So, applying $N$ times the optimal single-system
POVM is the best joint measurement when $N$ replicas are available
since it yields $\sigma_x^2(N)=N^{-1}\sigma_x^2(1)$ and
$\sigma_p^2(N)=N^{-1}\sigma_p^2(1)$. Hence, using
Eq.~(\ref{eq:holevobound}) for a coherent state ($\Delta
\hat{x}^2=\Delta  \hat{p}^2=1/2$) and requiring
$\sigma_x^2(N)=\sigma_p^2(N)$, the tightest bound is obtained for
$g_x=g_p$. It yields \bed \overline{\sigma}^2_{N,\infty}=1/N, \eed
which, combined with Eq.~(\ref{eq:addvar}), gives the minimum
noise variance induced by cloning, Eq.~(\ref{eq:mainresult}).

\subsection{Implementation of Gaussian QCMs}\label{sec:cvimpl}

Now that we have derived upper bounds on optimal cloning, we shall
show that these bounds are achievable, and exhibit explicit
optimal cloning transformations. Remarkably, these transformations
have a fairly simple implementation, when the quantum oscillator
corresponds to a light mode: it requires only a phase-insensitive
linear amplifier and a network of beam splitters\footnote{Note
that another implementation, with the same performances, involving
a circuit of C-NOT gates has also been proposed \cite{cer00d}.}.
We shall also discuss the link between the issue of optimal
quantum cloning and that of optimal amplification of quantum
states.

\subsubsection{Definitions and requirements}

Let us first state what we expect from a quantum cloning machine.
Let $\ket{\Psi}=\ket{\alpha}^{\otimes N} \otimes \ket{0}^{\otimes
M-N} \otimes \ket{0}_z$ denote the initial joint state of the $N$
input modes to be cloned (all prepared in the coherent state
$\ket{\alpha}$), the additional $M-N$ blank modes, and an
ancillary mode $z$. The blank modes and the ancilla are assumed to
be initially in the vacuum state $\ket{0}$. Let
$\{\hat{x}_k,\hat{p}_k\}$ denote the pair of quadrature operators
associated with each mode $k$ involved by the cloning
transformation\footnote{In what follows, we sometimes omit the
hats on operators when the context is clear.}, $k=0 \ldots N-1$
refers to the $N$ original input modes, and $k=N \ldots M-1$
refers to the additional blank modes. Cloning can be thought of as
some unitary transformation \bed U: \mathcal{H}^{\otimes M+1} \to
\mathcal{H}^{\otimes M+1}: \ket{\Psi} \to  U
\ket{\Psi}=\ket{\Psi''}. \eed Alternatively, in the Heisenberg
picture, this transformation can be described by a canonical
transformation of the operators $\{x_k,p_k\}$:
\ba\label{eq:clontsfheis} x''_k  =  U^{\dagger} \, x_k \, U&\;,\;&
p\,''_k =  U^{\dagger} \, p_k \,U\,. \ea We work in the Heisenberg
picture because cloning turns out to be much simpler to study from
that point of view. We now impose several requirements on the
transformation Eq.~(\ref{eq:clontsfheis}) that translate the
expected properties for an optimal cloning transformation.

First, we require the $M$ output modes quadratures have the same
mean values as the the input mode: \beqa
\langle {x''_k}\rangle=\bra{\psi}{x_0} \ket{\psi}, \quad k=0 \ldots M-1, \\
\langle {p\,''_k}\rangle=\bra{\psi}{p_0} \ket{\psi}, \quad k=0
\ldots M-1. \eeqa This means that the state of the clones is
centered on the original coherent state. Our second requirement is
covariance with respect to rotation in phase space. Coherent
states have the property that quadrature variances are left
invariant by complex rotations in phase space. That is, for any
mode $k$ involved in the cloning process and for any operator
$v_k=cx_k+dp_k$ (where $c,d$ are complex numbers satisfying
$|c|^2+|d|^2=1$), we have: \beq
\Delta v_k^2 = \moy{v_k^2}-\moy{v_k}^2=\textrm{vacuum fluct.}=1/2. \nonumber \\
\eeq We impose this property to be conserved through the cloning
process. Taking optimality into account,
Eq.~(\ref{eq:mainresult}), rotation covariance yields:
\beq\label{eq:expectnoise}
\sigma^2_{v''_k}=(\frac{1}{2}+\frac{1}{N}-\frac{1}{M}), \eeq where
$v''_k=cx''_k+dp''_k$.

The third requirement is, of course, the unitarity of the
transformation. In the Heisenberg picture, unitarity translates
into demanding that the commutation rules be conserved through the
evolution \cite{cav82}: \beq
[{x_j}'',{x_k}'']=[{p_j}'',{p_k}'']=0, \hspace{0.1cm}
[{x_j}'',{p_k}'']=i \delta_{jk}. \eeq

\subsubsection{Optimal Gaussian $1\rightarrow 2$ QCM}

Let us first focus on duplication ($N=1,M=2$). A simple
transformation meeting the three conditions mentioned above is
given by: \beqa \label{dupli-xp}
x_0''&=&x_0+\frac{x_1}{\sqrt{2}}+\frac{x_z}{\sqrt{2}}, \qquad
p_0''=p_0+\frac{p_1}{\sqrt{2}}-\frac{p_z}{\sqrt{2}}, \nonumber \\
x_1''&=&x_0-\frac{x_1}{\sqrt{2}}+\frac{x_z}{\sqrt{2}}, \qquad
p_1''=p_0-\frac{p_1}{\sqrt{2}}-\frac{p_z}{\sqrt{2}}, \nonumber \\
x_z'&=&x_0+\sqrt{2}\, x_z, \qquad \qquad ~ p_z'=-p_0+\sqrt{2}\,
p_z. \eeqa This transformation clearly conserves the commutation
rules, and yields the expected mean values ($\moy{x_0},\moy{p_0}$)
for the two clones (modes $0''$ and $1''$). One can also check
that the quadrature variances of both clones are equal to $1$, in
accordance with Eq.~(\ref{eq:expectnoise}). This transformation
actually coincides with the cloning machine introduced by
\textcite{cer00d}. Interestingly, we note here that the state in
which the ancilla $z$ is left after cloning is centered on
$(x_0,-p_0)$, that is the {\em phase-conjugated} state
$\ket{\bar{\alpha}}$. This means that, in analogy with the
universal qubit cloning machine \cite{buz96}, the
continuous-variable cloner generates an anti-clone (or
time-reversed state) together with the two clones.

\begin{figure}[h]
\begin{center}
\includegraphics[width=5.5cm,height=3.5cm]{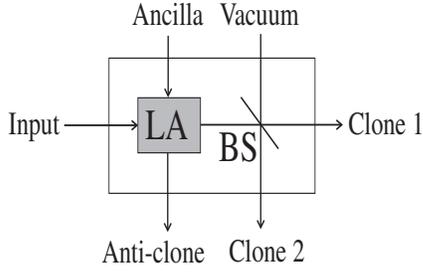}
\caption{Implementation of the optimal Gaussian $1 \to 2$ QCM for
light modes. LA stands for Linear Amplifier, and BS represents a
balanced Beam Splitter.}\label{fig:clon12}
\end{center}
\end{figure}

Now, let us show how this duplicator can be implemented in
practice. Eq.~(\ref{dupli-xp}) can be interpreted as a two-step
transformation: \beqa\label{dupli} a'_0 &=&  \sqrt{2} a_0 +
a_z^{\dagger}, \qquad
a'_z  =  a_0^{\dagger} + \sqrt{2} a_z, \nonumber\\
a''_0 &=&  \frac{1}{\sqrt{2}}(a'_0+a_1), \quad a''_1 =
\frac{1}{\sqrt{2}}(a'_0-a_1). \eeqa As shown in
Fig.~\ref{fig:clon12}, the interpretation of this transformation
is straightforward: the first step (which transforms $a_0$ and
$a_z$ into $a_0'$ and $a_z'$) is a phase-insensitive amplifier
whose (power) gain $G$ is equal to 2, while the second step (which
transforms $a_0'$ and $a_1$ into $a_0''$ and $a_1''$) is a
phase-free 50:50 beam splitter. Clearly, rotational covariance is
guaranteed here by the use of a {\em phase-insensitive} amplifier.
As discussed by \textcite{cav82}, the ancilla $z$ involved in
linear amplification can always be chosen such that $\moy{a_z}=0$,
so that we have $\moy{a''_0}=\moy{a''_1}=\moy{a_0}$ as required.
Finally, the optimality of our cloner can be confirmed from known
results on linear amplifiers. For an amplifier of gain $G$, the
quadrature variances of $a_z$ are bounded by \cite{cav82}:
\beq\label{amplibound} \sigma^2_{a_z} \geq (G-1)/2. \eeq Hence,
the optimal amplifier of gain $G=2$ yields $\sigma^2_{a_z} =1/2$,
so that our cloning transformation is optimal according to
Eq.~(\ref{eq:mainresult}).

\subsubsection{Optimal Gaussian $N\rightarrow M$ QCM}

\begin{figure}
\begin{center}
\includegraphics[width=7.5cm,height=4.5cm]{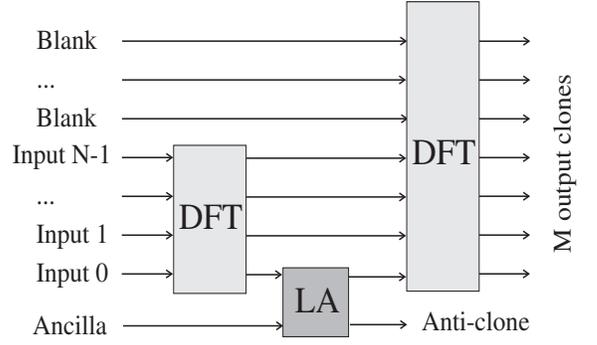}
\caption{Implementation of the optimal Gaussian $N \to M$ QCM for
light modes. LA stands for Linear Amplifier, DFT for Discrete
Fourier Transform.}\label{fig:clonnm}
\end{center}
\end{figure}

Let us now derive an $N \to M$ cloning transformation. To achieve
cloning, energy has to be brought to each of the $M-N$ blank modes
in order to drive them from the vacuum state to a state which has
the desired mean value. We shall again perform this operation with
the help of a linear amplifier. From Eq.~(\ref{amplibound}), we
see that the cloning induced noise essentially originates from the
amplification process, and grows with the gain of amplifier. So,
we shall preferably amplify as little as possible. Loosely
speaking, the cloning procedure should then be as follows: (i)
concentrate the N input modes into one single mode, which is then
amplified; (ii) symmetrically distribute the output of this
amplifier amongst the $M$ output modes. A convenient way to
achieve these concentration and distribution processes is provided
by the Discrete Fourier Transform (DFT). Cloning is then achieved
by the following three-step procedure (see Fig.~\ref{fig:clonnm}).
First step: a DFT (acting on $N$ modes), \beq
a'_k=\frac{1}{\sqrt{N}} \sum_{l=0}^{N-1} \exp(ikl 2\pi/N) \; a_l,
\eeq with $k=0\ldots N-1$. This operation concentrates the energy
of the $N$ input modes into one single mode (renamed $a_0$) and
leaves the remaining $N-1$ modes ($a'_1 \ldots a'_{N-1}$) in the
vacuum state. Second step: the mode $a_0$ is amplified with a
linear amplifier of gain $G=M/N$. This results in \beqa
a'_0&=&\sqrt{\frac{M}{N}} \; a_0 + \sqrt{\frac{M}{N}-1} \;
a_z^{\dagger},
\nonumber \\
a'_z&=&\sqrt{\frac{M}{N}-1} \; a_{0}^{\dagger}+\sqrt{\frac{M}{N}}
\; a_z. \eeqa Third step: amplitude distribution by performing a
DFT (acting on $M$ modes) between the mode $a'_0$ and $M-1$ modes
in the vacuum state: \beq a''_k=\frac{1}{\sqrt{M}}
\sum_{l=0}^{M-1} \exp(ikl 2\pi/M) \; a'_l, \eeq with $k=0\ldots
M-1$, and $a'_i=a_i$ for $i=N \ldots M-1$. The DFT now distributes
the energy contained in the output of the amplifier amongst the
$M$ output clones.

It is readily checked that this procedure meets our three
requirements, and is optimal provided that the amplifier is
optimal, that is $\sigma^2_{a_z}=[(M/N)-1]/2$. The quadrature
variances of the $M$ output modes coincide with
Eq.~(\ref{eq:mainresult}). As in the case of duplication, the
quality of cloning decreases as $\sigma^2_{a_z}$ increases, that
is \emph{ amplifying coherent states, or cloning them with the
same error for each clone, are two equivalent problems}. For $1
\to 2$ cloning, we have seen that the final amplitude distribution
amongst the output clones is achieved with a single beam splitter.
In fact, any unitary matrix such as the DFT used here can be
realized with a sequence of beam splitters and phase shifters
\cite{rec94}. This means that the $N\to M$ cloning transformation
can be implemented using only passive elements except for a single
linear amplifier. An explicit sequence of beam splitters achieving
a DFT on $M$ modes is given by \textcite{bra01sof}.

Finally, we note that, if {\em squeezed states} are put in rather
than coherent states, the transformations and circuits presented
here maintain optimum cloning fidelities, provided all auxiliary
vacuum modes (the blank modes and the ancillary mode $z$) are
correspondingly squeezed. This means, in particular, that the
amplifier mode $z$ needs to be controlled which requires a device
different from a simple phase-insensitive amplifier, namely a
two-mode parametric amplifier. One can say that the cloning
machine capable of optimum cloning of all squeezed states with
{\it fixed} and {\it known} squeezing then operates in a
non-universal fashion with respect to all possible squeezed states
at the input \cite{cer00c}.

\subsection{Other continuous variable QCMs}\label{sec:cvothers}

We conclude this Section by summarizing some interesting
developments in continuous variable cloning.

\emph{Other figures of merit.--} The universal Gaussian machines
presented in \ref{sect:opticlongauss} and \ref{sec:cvimpl} have
been derived requiring that the noise of the output clones be
minimum. But, one could have used other figures of merit to judge
the quality of the output clones. Then, would we have obtained
different solutions? Another related issue is: do we get better
cloners if the Gaussian assumption is relaxed? \textcite{cer04}
have proved that if one chooses the global
fidelity\footnote{Global fidelity was introduced in \ref{sswern}
for the case of discrete variables. Recall that in that case, the
optimization of the global and of the single-copy fidelity leads
to the same optimal UQCM.} as the figure of merit, then the
universal Gaussian cloner turns out to be optimal too. But
surprisingly, the Gaussian assumption is too restrictive if the
goal is to optimize the single-clone fidelity. For instance, for
$1 \to 2$ cloning, there exists a non-Gaussian operation whose
output clones have a fidelity of $0.6826$ with the original for
all coherent states, improving on the universal Gaussian machine,
which achieves a fidelity of $2/3 \approx 0.6666$, see
Eq.~(\ref{eq:fidelity}).

\emph{Optimal cloning for finite distributions of coherent
states.--} In devising optimal cloning machines, we require that
all coherent states be cloned with an equal quality. In other
words, we devised cloning machines which are optimal for a
distribution of coherent states in phase space which is flat. But,
for practical reasons, it is interesting to consider situations
where the coherent states to be cloned are produced according to a
\emph{finite} distribution over phase space --- in other words, to
drop the requirement of universality over all coherent states. In
particular, the case has been studied \cite{gro02th,coch04} in
which the coherent states to be cloned are produced according to a
Gaussian distribution \beq P(\alpha)=\frac{1}{2 \pi \Sigma^2}
e^{-|\alpha|^2/2\Sigma^2}\,. \eeq It is easily seen that in this
setting, the cloning procedures we have considered so far do not
produce clones with optimal fidelities. For instance, if
$P(\alpha)$ is a sufficiently peaked distribution, then a very
trivial cloning machine, from which the first output clone is the
unaffected original and the second clone is a mode prepared in the
vacuum state, already achieves better fidelities than the
universal Gaussian cloner. Actually, one can prove that for all
values of $\Sigma$, there is a cloning machine achieving a
single-clone fidelity of \beq F= \left\{
\begin{array}{ll}
\frac{4 \Sigma^2+2}{6 \Sigma^2+1}, \hspace{0.5cm} \Sigma^2 \geq \frac{1}{2}+\frac{1}{\sqrt{2}},  \\
 \frac{1}{(3-2\sqrt{2}) \Sigma^2+1},  \hspace{0.5cm} \Sigma^2 \leq \frac{1}{2}+\frac{1}{\sqrt{2}}. \\
\end{array}
\right. \eeq Interestingly, such a cloning machine can be achieved
using the setup shown in Fig.~\ref{fig:clon12}, but where the gain
is adapted to the distribution of coherent states: \bed G=\frac{8
\Sigma^4}{(2 \Sigma^2+1)^2}. \eed

\section{Application of quantum cloning to attacks in quantum cryptography}
\label{sec4}

\subsection{Generalities}

As stated in the introduction, the relationship between the
no-cloning theorem and the security of quantum cryptography was
already pointed out in the first protocol \cite{ben84}. Let us
briefly sketch here the common structure behind any protocol; we
refer the reader to the review by \textcite{gis02} for a more
thorough view on quantum cryptography. A sender, Alice, encodes
some classical information on a quantum state, chosen among a set
of non-orthogonal alternatives $\{\ket{\psi_i}_A\}$. The
so-prepared particle goes to a receiver, Bob, who randomly chooses
a measurement from a pre-established set of measurements. When
more than two states are used for the encoding, the exchange of
the particles is usually followed by a "sifting" phase, in which
Alice reveals something of the encoding (e.g. the basis to which
each state belongs), allowing Bob to know if he has done the good
measurement. After this process, Alice and Bob share a list of
classically correlated symbols. Using well-established techniques
from classical information theory, this list can be transformed
into a secret key\footnote{Actually, when discrete-level quantum
states are used, one normally tailors the protocol in such a way
that, in the absence of an eavesdropper and of errors, the
correlation between Alice and Bob is perfect without any classical
processing, i.e. it already constitutes a perfect secret key.
However, this can no longer be done for protocols using continuous
variables.}, which is later consumed for sending private
information by means of the one-time pad. So quantum cryptography
is actually quantum key distribution (QKD).

Although everything that takes place in Alice and Bob's sites is
secure\footnote{This is a very reasonable assumption for any
cryptographic scenario. Indeed, it seems difficult to design a
secure protocol if one cannot exclude the possibility that Eve has
access to Alice's preparation of quantum states, or Bob's
measurement results.}, this is no longer the case for the channel
connecting them. This means that an eavesdropper, usually called
Eve, can freely interact with the states while they propagate and
try to extract information. Eve is allowed to perform the most
general action consistent with quantum mechanics. In particular
then, she is {\em limited by the no-cloning theorem}: contrary to
what happens for classical information that can be amplified at
will, when Eve obtains information on the state sent by Alice, the
state used for the encoding is perturbed and she introduces
errors. The larger the information obtained by Eve is, the more
the state is perturbed, and consequently the larger is the error
rate in the correlations between Alice and Bob.

In fact, QKD is secure because one of the following cases happens:
{\em either} the error rate observed by Alice and Bob is lower
than a critical value $D_c$, in which case a secret key can be
extracted using techniques of classical information theory; {\em
or} the error rate is larger than $D_c$, in which case Alice and
Bob throw their data away and never use them to encode any
message. In other words, the eavesdropper can either lose the game
or prevent any communication, but will never gain any information.

All this reasoning is nice, provided that Alice and Bob are able
to find the value of the threshold $D_c$ for the protocol that
they want to use. That's why it is important to establish {\em
quantitative} trade-offs between the information acquired by Eve
and the error rate. For this calculation, one should assume that
Eve has applied the most powerful strategy consistent with quantum
mechanics. Therefore, the problem of estimating Eve's information
for a given disturbance is equivalent to finding her optimal
eavesdropping attack on the protocol that is used. This is a very
difficult problem and, to date, the complete solution is not known
for any of the existing protocols. Nevertheless, the problem can
be solved if Eve is restricted to the so-called {\em incoherent
attacks}. In what follows, we mainly focus on these attacks, that
involve QCMs. The last paragraph of this subsection, however, will
be devoted to the possibility of more general quantitative links
between QKD and cloning.

\subsection{Incoherent attacks and QCMs}

\subsubsection{Generalities}

An incoherent attack is defined by two conditions: (i) Eve
interacts individually and in the same way with the states
travelling from Alice to Bob; (ii) she measures the quantum
systems she has kept after the possible sifting
phase\footnote{Therefore, individual attacks require a quantum
memory.} but \emph{before} any reconciliation process has started.
In other words, the hypothesis is that after the sifting phase,
Alice, Bob and Eve share a list of classical random variables,
identically distributed according to a probability law $P(A,B,E)$.
Under this hypothesis, the fraction of secret bits $R$ that can be
extracted by Alice and Bob using reconciliation protocols with
one-way communication satisfies the bound of \textcite{csi78}
\ba\label{ckbound} R&=&I(A:B)\,-\,\min \big\{I(A:E),I(B:E)\big\}
\ea where $I(X:Y)=H(X)+H(Y)-H(XY)$ is the mutual information
between two parties\footnote{The function $H$ is the usual Shannon
entropy. While we were finishing this review, the idea of
"pre-processing" was introduced in quantum cryptography
\cite{kra04,ren05}. Quite astonishingly, these authors found that
security bounds can be improved by letting Alice randomly flip
some of her bits. The reason is that this procedure decreases
Alice's correlations with Eve much more than her correlations with
Bob. This result implies that, apart from the six-state protocol,
in which the attacks depend on a single parameter which is the
quantum bit error rate (QBER), the truly optimal incoherent
attacks may not be those which have been presented in the previous
literature. Since this is an open research problem, we haven't
taken these new considerations into account in the main text.}.
This result formalizes the intuition according to which, if Eve
has "as much information as Alice and Bob", it is impossible to
extract a secret key. Therefore, Eve's optimal individual attack
is the one that, for a given error rate $D$ --- that is, for a
given value of $I(A:B)=1-H(D)$ --- maximizes $I(A:E)$ and
$I(B:E)$. This defines the figure of merit for eavesdropping with
incoherent attacks.

If we go back to the physical implementation of such attacks, we
see that Eve is going to "transfer" some information about the
original state onto the state of a particle that she keeps and
measures later. Under this perspective, it seems rather natural to
guess that the interaction defining the best individual attack is
the $1\rightarrow 2$ asymmetric QCM that clones in an optimal way
all possible preparations by Alice, i.e. the set of states
$\{\ket{\psi_i}_A\}$; although there is no {\em a priori} link
between the optimality of cloning, based on the single-copy
fidelity (\ref{fiddef}), and the optimality of eavesdropping
defined just above. This intuition was proved to hold in the
following cases:
\begin{itemize}
\item BB84 protocol \cite{ben84}: Alice chooses her preparation
among the four states $\ket{\pm x}$ and $\ket{\pm y}$ of a qubit,
which belong to the equator of the Bloch sphere. One can see that
Eve's optimal individual attack uses the asymmetric
phase-covariant cloning machine (see \ref{ssstate}) \cite{fuc97}.
The security condition assumes an easy form: a secret key can no
longer be extracted as soon as $F_{Bob}=F_{Eve}$, that is, the
critical value for the error rate is $D_c^{incoh}=1-F_{Eve}$.
Using (\ref{fidxy}), $D_c^{incoh}\simeq 14.6\%$. We shall come
back to this example in full detail in the next paragraph.

\item Six-state protocol \cite{bru98c,bec99}: Alice's preparation
is the same as in BB84 plus the poles of the sphere, $\ket{\pm
z}$. Again, the interaction defining the optimal attack is the
universal asymmetric QCM (see \ref{ssasym}). Indeed, it turns out
that to optimally clone these six states is equivalent to clone
all the states in the Bloch sphere. One can see that the critical
disturbance such that $R$ (\ref{ckbound}) goes to zero is
$D_c^{incoh}\simeq 15.7\%$. Note that in the six-state protocol
Eve's attack is more limited than in BB84 because she has to
(imperfectly) clone all the states in the sphere. This intuitively
explains why Alice and Bob can tolerate a larger disturbance.

\item Continuous variable protocols, using both squeezed and
coherent states \cite{gro02,cer01}: there also exists a link
between security and the no-cloning theorem. Indeed, the
well-known security limit of 3 dB, common to all these protocols
for the case of direct reconciliation, can be understood as the
point where Eve's clone becomes equal to Bob's.
\end{itemize}

The connection between cloning machines and eavesdropping attacks
has also been exploited for other protocols and scenarios. For
instance, asymmetric $2\rightarrow 2+1$ cloning machines have been
discussed for eavesdropping on practical implementations of QKD,
with no claim of optimality \cite{aci04b,cur04,nie04}. Going to
higher dimensional systems, the relation between cloning machines
and incoherent eavesdropping strategies has been analyzed
\cite{bru02,cer02a}. Here, optimality is conjectured but not
proved (see in this context \textcite{kas04}). In the case of the
protocol invented by \textcite{sarg} (SARG04), the optimal
incoherent eavesdropping is not known, but the best attack which
has been found by \textcite{bra05} does not make use of the
corresponding optimal cloner (which would be the phase-covariant
one, as for BB84).

\subsubsection{Optimal incoherent attack on the BB84 protocol}
\label{sssopt}

As a completely worked-out example, we describe the optimal
incoherent attack on the BB84 protocol. Suppose that the BB84
protocol is run with the bases of the eigenstates of $\si_x$ and
$\si_y$; it is then no surprise that the optimal incoherent attack
is obtained when Eve makes a copy of each qubit using the
phase-covariant cloner described in \ref{ssstate} \cite{fuc97}.
However, for eavesdropping in cryptography there is a difference
between the two implementations that we presented, the one
(\ref{uning}) without ancilla and the one (\ref{unicerf}) with an
ancillary qubit \cite{dur03b,aci04a,aci04b}. Intuitively, the
reason is that some kind of information is stored in the ancilla
as well, and Eve has an access to it. Here we show in detail what
happens.

Let's study the {\em cloner without ancilla} first. For
simplicity, we focus on an item where Alice has sent $\ket{+x}$,
and suppose that Bob has measured $\si_x$ so that the item will be
kept after the bases-reconciliation. Using (\ref{uning}), the
flying qubit becomes entangled to Eve's qubit according to \ba
\ket{\Gamma}_{BE}&=&\frac{1}{\sqrt{2}}\,\big(\ket{0}\ket{0}+\cos\eta\ket{1}\ket{0}
+ \sin\eta \ket{0}\ket{1}\big)\,. \label{entgstate}\ea Bob's qubit
is thus in the state
$\rho_B=\demi(\one+\sin^2\eta\si_z+\cos\eta\si_x)$, so that the
measurement of $\si_x$ gave him the correct outcome with the
probability $F_{AB}=\sandwich{+x}{\rho_B}{+x}=
\frac{1}{2}(1+\cos\eta)$, which is indeed the fidelity of his
clone as expected. The Alice-Bob mutual information is therefore
$I(A:B)=1-H(F_{AB})$. Similarly, Eve's qubit is in the state
$\rho_E=\demi(\one+\cos^2\eta\si_z+\sin\eta\si_x)$, whence she
will guess the state sent by Alice correctly with probability
$F_{AE}= \frac{1}{2}(1+\sin\eta)$. The Alice-Eve mutual
information is therefore $I(A:E)=1-H(F_{AE})$. Obviously,
$I(A:B)=I(A:E)$ for $\eta=\frac{\pi}{4}$, that is for an error
rate $D_{AB}=1-F_{AB}\simeq 0.1464$. However, the security
criterion for one-way communication (\ref{ckbound}) says that
$I(A:B)$ must be larger than the {\em minimum} between $I(A:E)$
and $I(B:E)$, so we need to compute the mutual information Bob-Eve
as well. From the state (\ref{entgstate}), we can compute the
probability that Eve's symbol is equal to Bob's, knowing that both
measure $\si_x$: $P_{BE}=|\braket{+x,+x}{\Psi_{BE}}|^2+
|\braket{-x,-x}{\Psi_{BE}}|^2=\demi\left(1+\demi \sin
2\eta\right)$; whence the mutual information $I(B:E)=1-H(P_{BE})$.
It can be verified that $I(B:E)<I(A:B)$ whenever $D_{AB}<\demi$;
from Eq.~(\ref{ckbound}), Alice and Bob could always extract a
key, as long as their correlation is not zero\footnote{We note
here that this analysis was done first in the Introduction of the
article by \textcite{sca01}; but an unfortunate mistake in the
computation of $P_{BE}$, Eq.~(5), prevented them from reaching the
correct conclusion.}. This is too good to be true; and indeed, the
use of the machine with an ancillary qubit yields a more
reasonable scenario.

To study the {\em machine with an ancilla}, we suppose that Alice
and Bob use the same basis, but we consider both eigenstates of
$\si_x$. Using (\ref{unicerf}), the flying qubit $\ket{\pm x}_{A}$
becomes entangled to Eve's two qubits according to \ba
\ket{\Gamma^{\pm}}_{BE_1E_2}&=&\frac{1}{2}\,\big(\ket{000}+c\ket{011}
+ s\ket{101}\nonumber\\ && \pm c\ket{100} \pm s\ket{010} \pm
\ket{111}\big) \label{entgstate2}\ea where $c=\cos\eta$ and
$s=\sin\eta$. Bob's qubit is in the state
$\rho_B=\demi(\one\pm\cos\eta\si_x)$, which gives the same
fidelity as above as expected. The easiest way to see what Eve can
do with her two qubits consists of writing
$\ket{\Gamma^{\pm}}_{BE_1E_2}$ using the basis $\ket{\pm x}$ for B
and the Bell basis
$\ket{\Phi^{\pm}}=\frac{1}{\sqrt{2}}(\ket{00}\pm\ket{11})$,
$\ket{\Psi^{\pm}}=\frac{1}{\sqrt{2}}(\ket{01}\pm\ket{10})$ for
Eve's qubits, then in applying on Eve's qubits the unitary
transformation $\ket{\Phi^+}\rightarrow\ket{00}$,
$\ket{\Psi^+}\rightarrow\ket{10}$,
$\ket{\Phi^-}\rightarrow\ket{11}$,
$\ket{\Psi^-}\rightarrow\ket{01}$. After this transformation, the
states read \ba \ket{\tilde{\Gamma}^{\pm}}&=&\sqrt{F}\ket{\pm
x}\,\ket{\chi_{\pm}}\, \ket{0}\, \mp\, \sqrt{D}\ket{\mp x}\,
\ket{\chi_{\mp}}\,\ket{1} \label{gammas}\ea where
$F=\frac{1+c}{2}$ is Bob's qubit fidelity, $D=1-F$ the
disturbance, and $\ket{\chi_{\pm}}=\sqrt{F}\ket{0}\pm
\sqrt{D}\ket{1}$. Now Eve's strategy is clear. First, she measures
$\si_z$ on qubit $E_2$: if she finds $\ket{0}$, resp. $\ket{1}$,
she knows that Bob's bit is identical or opposite, respectively,
to Alice's bit. This information is deterministic, and implies
that Eve has as much information on Bob's bit as she has on
Alice's bit: $I(A:E)=I(B:E)$. This solves the main problem of the
machine without ancilla. For completeness, let's conclude the
calculation by computing $I(A:E)$. To guess Alice's bit, Eve must
distinguish between the two non-orthogonal states
$\ket{\chi_{\pm}}$ of qubit $E_1$, with {\em a priori}
probabilities $p_+=p_-=\demi$ since Alice sends $\ket{+x}$ and
$\ket{-x}$ with the same probability. It is known \cite{hel76}
that the maximal information she can obtain is $I(A:E)=1-H(P)$
where $P=\demi\big(1+\sqrt{1-|\braket{\psi_1}{\psi_2}|^2}\big)$.
Since $|\braket{\chi_+}{\chi_-}|=\cos\eta$, we recover the
expected result $P=\demi(1+\sin\eta)$.

In summary: without ancilla, Eve can make the best possible guess
on the bit sent by Alice (because the machine realizes the optimal
phase-covariant cloning) but has very poor information about the
result obtained by Bob. Adding the ancilla does not modify the
estimation of Alice's bit but allows Eve to deterministically
symmetrize her information on Alice and Bob's symbols. However,
the two machines are equally good from the point of view of
cloning.

\subsection{Beyond incoherent attacks}

All the links that we have discussed between QKD and quantum
cloning hold in the case of incoherent attacks. However, one can
also expect a relation between cloning and eavesdropping in more
general security analysis. Consider the BB84 protocol, and assume
that Eve interacts individually with the states sent by Alice, but
she can delay her measurement until the end of the reconciliation
process and then possibly perform collective measurements. These
types of attacks are often called {\em collective}. The results of
\textcite{dev03} and of \textcite{ren04} imply that there exists a
protocol achieving a key rate \ba\label{dwbound}
    R&=&I(A:B)\,-\,\min \big\{\chi(A:E_Q),\chi(B:E_Q)\big\} ,
\ea which can be understood as the generalization of
(\ref{ckbound}) to the case where Eve's variables are quantum
(whence the index $Q$). The quantity $\chi$ is the so-called
Holevo bound \cite{hol73}, which bounds the maximal information on
Alice or Bob's symbol accessible to Eve through her quantum
states. Indeed, the presence of Eve's attack defines an effective
channel between Alice (or Bob) and Eve. For this channel, when
Alice encodes the symbol $X=0,1$ on the quantum state
$\ket{\psi_X}$, Eve receives the state $\rho_E^X$ obtained by
tracing out the qubit that goes to Bob. Holevo's bound then reads
\begin{equation}\label{holevo}
    \chi(X:E)=S(\rho_E)-\frac{1}{2}S(\rho_E^0)-\frac{1}{2}S(\rho_E^1),
\end{equation}
where $S$ denotes the von Neumann entropy and
$\rho_E=(\rho_E^0+\rho_E^1)/2$.

If Eve uses the phase-covariant cloning machine, we know
$\rho_E^{0,1}=\mbox{Tr}_B
\left(\ket{\Gamma^{\pm}}\bra{\Gamma^{\pm}}\right)$ from
Eq.~(\ref{gammas}) and can compute $R$. After some simple algebra,
one can see that the critical error $D_c$ at which $R$
(\ref{dwbound}) is zero is defined by $1-2H(D_c)=0$. Remarkably,
this equation is the same as in the \textcite{sho00} proof of
security of the BB84 protocol, and leads to a critical disturbance
of $D_c\approx 11\%$. The Shor-Preskill proof of security does not
make any assumption on Eve's attack: it is thus remarkable that
the same bound can be reached by a collective attack in which the
individual quantum interaction is defined by the phase-covariant
QCM. Actually, the attack based on the phase-covariant cloning
machine is optimal, in the sense that it minimizes (\ref{dwbound})
for a fixed disturbance.

\subsection{Conclusive balance}

The relation between the no-cloning theorem and the security of
quantum cryptography is certainly deep. However, it is not clear
at all how to associate {\em quantitative} results for
cryptography to some explicit form of imperfect cloning, because
cloning is not equivalent to eavesdropping. In particular, the
relevant figure of merit that define Eve's optimal attack is {\em
a priori} unrelated to the single-copy fidelity (\ref{fiddef})
that is optimized when constructing QCMs --- see also the
discussion by \textcite{bru98a}. Still, the connection has proved
to be strong and fruitful in the case of individual attacks, and
possibly even beyond.

\section{Stimulated emission as optimal cloning of discrete variables in optics}
\label{sec5}

In this Section, we discuss how the well-known amplification
phenomenon of stimulated and spontaneous emission of light is
closely related to optimal universal cloning. The results are
stated and commented on in \ref{ssclampl}; in \ref{ssphenom}, we
re-derive the main results using a phenomenological model.

\subsection{Cloning as amplification}
\label{ssclampl}

\subsubsection{Encoding of discrete states in different modes}

Section \ref{sec3} was devoted to the cloning of coherent states
of a quantum oscillator; all the discussion, especially about the
implementations, was carried out having in mind a single mode of
the light field as an example of a quantum oscillator. In this
Section, we consider the light field too, but in a different
perspective: the quantum system is now the discrete-level system
encoded in some modes of the field. The typical example here is
polarization: for a given energy $\omega$, the light field has two
independent modes $a_H(\omega)$ and $a_V(\omega)$, corresponding
to two orthogonal polarizations "horizontal" and "vertical". We
can then define a qubit as \ba
a_H^{\dagger}(\omega)\ket{\textrm{vac}}&=&\ket{\mathbf{0}}\\
a_V^{\dagger}(\omega)\ket{\textrm{vac}}&=&\ket{\mathbf{1}} \ea
where $\ket{vac}$ is the vacuum state of the field. According to
this correspondence, for any pair of complex numbers $c_H,c_V$
such that $|c_H|^2+|c_V|^2=1$, we can define a creation operator
$a^{\dagger}_\psi=[c_Ha_H^{\dagger}(\omega)+c_V
a_V^{\dagger}(\omega)]$ such that \ba
a^{\dagger}_\psi\ket{vac}&=&\ket{\psi}\,=\,c_H\ket{\mathbf{0}} +
c_V\ket{\mathbf{1}}\,.\ea In this sense, polarization in a given
frequency mode defines a qubit. Obviously, the N-photon Fock state
in which all the photons are prepared in the state $\ket{\psi}$
reads \ba \ket{\psi}^{\otimes
N}&=&\frac{\left(a^{\dagger}_\psi\right)^N}{\sqrt{N!}}\,\ket{\mbox{vac}}\,.
\ea The construction clearly generalizes: we can encode a qudit
with any $d$ orthogonal modes $a_1,...,a_d$.

In Section \ref{sec3} the unknown parameters were the parameters
defining a coherent state in a given mode (i.e., a quantum
continuous variable); here, a Fock state of $N$ photons is
prepared in a mode $a_\psi$ which is a linear combination of $d$
modes $a_j$: the unknown parameters are the coefficients of the
linear combination (that is, a qudit). Therefore, we are going to
refer back to the cloning of discrete-level systems (Section
\ref{sec2}). In all that follows, for simplicity, we discuss
explicitly the example of polarization in a single energy mode. As
might be expected, the results extend to any discrete level system
encoded in field modes \cite{fan02}; we sketch it in paragraph
\ref{sssfiber} for the case of time-bin encoding.

\subsubsection{Main result}

Consider a light amplification process based on stimulated
emission. We consider two orthogonal polarization modes of a
monochromatic component of the field, and suppose that (i) $N$
photons of a given {\em unknown} polarization are already present
in the medium, and (ii) the component of the field associated to
exactly $M>N$ photons is post-selected after amplification.
Because spontaneous emission is always present, it is impossible
that all $M$ photons are deterministically emitted in the same
polarization mode as the input ones: even for large $N$, there
will always be a small probability that a photon is emitted in the
orthogonal mode. The claim is that, if the probabilities of
emission are independent of the polarization, this amplification
process attains the optimal fidelity for universal $N\rightarrow
M$ cloning of qubits. This was noticed in the very early days of
quantum cloning \cite{woo82,mil82,man83} for the $1\rightarrow 2$
process, and was generalized more recently to any cloning process
\cite{sim00,kem00}. In the rest of this Section, we derive the
same results using the more phenomenological approach sketched in
\textcite{fas02}.

\subsection{Phenomenological model}
\label{ssphenom}

\subsubsection{Definition and fidelity $1\rightarrow 2$}

Consider an inverted medium that can emit photons of any
polarization with the same probability (thus, we introduce by hand
the assumption of universality). We focus on a monochromatic
component of the field. Suppose that $N$ photons are initially
present in a given polarization mode, say $\ket{V}$; and suppose
that at the output of the amplifier $M=N+k$ photons are found, the
initial ones plus $k$ new ones that have been emitted by the
medium\footnote{Photons are bosons: the output state will be a
{\em symmetrized} state of the $M$ photons in which $N$ photons
are certainly in state $\ket{V}$ and the other ones are in a
suitable state. So it does not really make sense to speak of the
"initial photons" as if they had conserved any distinctive
property whatsoever after amplification. Still, one can use this
loose language, provided that the relation between spontaneous and
stimulated emission, Eq.~(\ref{factorial}), is assumed. This
relation is a consequence of the bosonic nature of the field.}.
For this amplification process, the {\em single-copy fidelity} is
the probability of an output photon picked at random to be
polarized as the input ones. The no-cloning theorem tells us that
the $k$ additional photons cannot be deterministically in the same
polarization mode $\ket{V}$ as the $N$ input ones; and indeed, we
know that stimulated emission is always associated with
spontaneous emission.

The derivation of the fidelity for an amplification $1\rightarrow
2$ can be easily described. If a photon is present in mode
$\ket{V}$, a second photon in this mode can be emitted either by
spontaneous or by stimulated emission, the two processes being
equiprobable; while a photon in mode $\ket{H}$ can be emitted only
by spontaneous emission. Thus the probabilities $P[2,0|1,0]$ and
$P[1,1|1,0]$ that the new photon is emitted in the same mode as
the input or in the orthogonal mode, are related to one another as
$P[2,0|1,0]=2\,P[1,1|1,0]$: the probability for the new photon to
be polarized along $\ket{V}$ is $\frac{2}{3}$. If we now pick a
photon out of the two, with probability $\demi$ it is the original
one, whose polarization is certainly $\ket{V}$; with probability
$\demi$, it is the new one. So, the probability for finding one of
the output photons in mode $\ket{V}$ (the fidelity) is
$\demi\times 1+\demi\times \frac{2}{3}=\frac{5}{6}$, exactly the
same as for optimal symmetric universal $1\rightarrow 2$ cloning.

In the rest of this Section, we generalize the same considerations
to derive the fidelity for the $N\rightarrow M$ cloning.

\subsubsection{Statistics of stimulated emission}

As a preliminary for what follows, we need to give the statistics
of the process of stimulated and spontaneous emission. This
amplification process will be completely described by the
probabilities $P[N+l,k-l|N,0]$, $0\leq l\leq k$, that $l$ photons
are emitted in mode $\ket{V}$ and $k-l$ in mode $\ket{H}$. We
normalize these probabilities so that they sum up to the total
probability of the process: \ba
\sum_{l=0}^{k}P[N+l,k-l|N,0]&=&P(N\rightarrow M)\,.
\label{sumprob}\ea We stated above the simplest example,
$P[2,0|1,0]=2\,P[1,1|1,0]$; now we want to show that the general
expression is \ba P[N+l,k-l|N,0]&=
&\frac{(N+l)!}{N!\,l!}\,P[N,k|N,0]\,. \label{factorial} \ea For
definiteness, we consider a medium formed of $\cal{N}$ "Lambda"
atoms, in which a unique excited state $\ket{e}$ can decay into
two orthogonal ground states $\ket{g_H}$ and $\ket{g_V}$ through
the emission of the correspondingly polarized photon. Omitting
coupling constants, the Hamiltonian describing the interaction
between the medium and the field is \ba
H&=&\sum_{j=1}^{{\cal{N}}}\left(a^{\dagger}_H\,\sigma^-_{H,j}+
a^{\dagger}_V\,\sigma^-_{V,j}\right)\,+\,\mbox{adj.} \ea where
$\si^-_{H,j}=\ket{g_H}\bra{e}$ and $\si^-_{V,j}=\ket{g_V}\bra{e}$
acting on atom $j$. The system is prepared so that all the atoms
are in the excited state, $N$ photons are in mode $\ket{V}$ and
none in mode $\ket{H}$: $\ket{\mbox{in}}=\ket{N,0\,;\,e,e,...,e}$.
The interaction leads to
$\ket{\psi(\tau)}=e^{-iH\tau/\hbar}\ket{\mbox{in}}$ where $\tau$
is the interaction time.

At the output, we post-select on the states such that exactly $k$
photons have been emitted; more specifically, we want $l$
additional photons in mode $\ket{V}$ and $k-l$ in mode $\ket{H}$.
By reading the state of the atoms after the interaction, one could
in principle know which atom has emitted which photon, so all the
possible output states are distinguishable. Consider all the
possible processes in which the first $k$ atoms have emitted a
photon --- all the other processes contribute with equal weight:
$\ket{\mbox{out}(\xi)}=\ket{N+l,k-l;
g_{\xi(1)},...,g_{\xi(k)},e,...,e}$ where $\xi$ is a $k$-item
sequence containing $l$ times the symbol "V" and $k-l$ times the
symbol "H". Then the probability $P[N+l,k-l|N,0]$ is proportional
to $\sum_{\xi}\left|
\sandwich{out(\xi)}{H^k}{in}\right|^2=\left(\begin{array}{c}k\\
l\end{array}\right)\,
\left|\sandwich{N+l,k-l}{{a^{\dagger}_V}^l{a^{\dagger}_H}^{k-l}}{n,0}\right|^2
\,=\,k!\,\frac{(N+l)!}{N!\,l!}$. This proves
Eq.~(\ref{factorial}). As a consequence of it, Eq.~(\ref{sumprob})
becomes \ba P(N\rightarrow
M)&=&P[N,k|N,0]\,\frac{(N+k+1)!}{(N+1)!\,k!} \label{sumprob2}\ea
since $\sum_{l=0}^{k} \frac{(N+l)!}{N!\,l!}\,
=\,\frac{(N+k+1)!}{(N+1)!\,k!}$. We can now go back to cloning and
prove the main result of this Section.

\subsubsection{Fidelity $N\rightarrow M$}

The fidelity of the amplification process is defined as usual, as
the probability of finding a photon in the same mode as those of
the input: \ba F_{N\rightarrow
M}&=&\frac{N+\moy{l}_{NM}}{M}\label{fidenm}\ea where \ba
\moy{l}_{NM}&=&\sum_{l=0}^{k}l\,
\frac{P[N+l,k-l|N,0]}{P(N\rightarrow M)}\label{moyl}\ea is the
average number of additional photons produced in the same mode as
the input. Inserting (\ref{factorial}) and (\ref{sumprob2}) into
(\ref{moyl}) and using $\sum_{l=0}^{k}l\,\frac{(N+l)!}{N!\,l!}
=\sum_{m=0}^{k-1}\,\frac{(N+1+m)!}{N!\,m!}$, we obtain
$\moy{l}_{NM}\,=\, k\,\frac{N+1}{N+2}$. Replacing $k=M-N$ we
obtain $F_{N\rightarrow
M}\,=\,\frac{1}{M}\big(N+(N-M)\frac{N+1}{N+2}\big)$ which is
exactly the Gisin-Massar result (\ref{fqubits}).

Our phenomenological model shows that the link between
amplification by an inverted medium and quantum cloning is
"semi-classical", in the following sense: the relation
(\ref{factorial}) is derived rigorously from quantum mechanics
(the bosonic nature of the field); but once this relation is
admitted, the rest becomes just classical event counting. Note in
particular how, due to Eq.~(\ref{factorial}), $\moy{l}_{NM}$ and
consequently $F_{N\rightarrow M}$ become independent of both
$P[N,k|N,0]$ and $P(N\rightarrow M)$. These last probabilities,
i.e. how frequent the process $N\rightarrow M$ is, are in general
difficult to compute and depend on the detailed physics of the
inverted medium (\textcite{sim00} and \textcite{kem00} provide
some examples). However, we know that whenever such an
amplification process takes place, it realizes the optimal
symmetric $N\rightarrow M$ UQCM.

\section{Experimental demonstrations and proposals}
\label{sec6}

This Section reviews the experiments that have been proposed and
often performed to demonstrate quantum cloning. They all refer to
{\em universal} cloning, symmetric or asymmetric. Phase-covariant
cloning has also been the object of recent proposals
\cite{fiu03,dec04}.

\subsection{Polarization of photons}

The connection between stimulated emission and quantum cloning
(Section \ref{sec5}) is the essential ingredient in most of the
optical implementations of qubit $1\rightarrow 2$ cloning
machines. The usual scheme consists of sending a single photon
into an amplifying medium. In the absence of this photon, the
medium will spontaneously emit photons of any polarization (or
mode). But if the photon is present, it stimulates the emission of
another photon in the same mode, i.e. this mode is enhanced.
However, the process of spontaneous emission can never be
suppressed, which means that the quality of the amplification
process is never perfect. This is indeed a manifestation of the
no-cloning theorem; remarkably, as discussed in detail in the
previous Section, it achieves {\em optimal} cloning.

Before discussing this kind of cloning, we must mention that one
of the first optical experiments that implemented the
Bu\v{z}ek-Hillery cloning (\ref{ssbh}) was an experiment using
only {\em linear optics} \cite{hua01}. The idea there is to
realize the three needed qubits with a single photon: one qubit is
the polarization, the other two are defined by the location of the
photon into four possible paths. As well-known, the optical device
called {\em polarizing beam-splitter} (PBS) realizes the CNOT gate
between he polarization and the path mode. The experimental setup
to achieve cloning is a suitable arrangement of PBS and optical
rotators. This being mentioned, we focus on cloning through
amplification processes.

\subsubsection{Experiments with parametric down-conversion}

\begin{center}
\begin{figure}
\includegraphics[width=8cm]{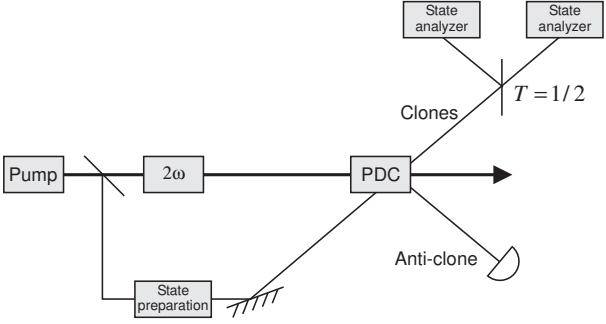}
\caption{Experimental
implementation of the $1\rightarrow 2$ universal cloning machine
for qubits. The laser works in pulsed mode. A very small fraction
of the pulse, below single-photon level, serves as probabilistic
preparation of the photon to clone. The rest of the pump pulse is
frequency doubled and sent to a non-linear crystal. If the
prepared photon is indistinguishable from one of the two
down-converted photons, optimal cloning is achieved. The other
down-converted photon is often called anti-clone.}
\label{figexpcl}
\end{figure}
\end{center}

Most optical implementations of the $1\rightarrow 2$ cloning
machine \cite{dem00,lam02,dem02} use {\em parametric
down-conversion} (PDC) as the amplification phenomenon (see
Fig.~\ref{figexpcl}). A strong laser pulse pumps a non-linear
crystal. With small probability the pulse is split into two
photons, called signal $S$ and idler $I$. For pulsed type-II
frequency degenerated PDC the Hamiltonian reads
\begin{equation}\label{PDCham}
    H=\gamma(a_{VS}^\dagger a_{HI}^\dagger-
    a_{HS}^\dagger a_{VI}^\dagger) + \textrm{ h.\,c. } .
\end{equation}
Notice that this Hamiltonian is invariant under the same unitary
operation in both polarization modes, $(VS,HS)$ and $(VI,HI)$. The
photon to be cloned and the the pump pulse propagate through the
crystal at the same time. Because of the Hamiltonian symmetry, one
can take as the state to clone,
$\ket{1,0}_S=a_{VS}^\dagger\ket{\mbox{vac}}$, without losing
generality. Indeed the rotational symmetry of the Hamiltonian
guarantees the covariance of the transformation. The state after
the crystal is
\begin{equation}\label{PDCout}
    \ket{\psi_{out}}=e^{-iHt}a_{VS}^\dagger\ket{\mbox{vac}} .
\end{equation}
We can expand the previous expression into a Taylor series. Since
the down-conversion process only happens with small probability,
we restrict our considerations to the first terms in the
expansion. The zero-order term simply corresponds to the case
where no pair of photons is produced, so at the output one finds
the initial state unchanged. The first order term is more
interesting, since the resulting normalized state gives
\begin{equation}\label{PDCout12}
    \ket{\psi_{1\rightarrow 2}}=\sqrt{\frac{2}{3}}\ket{2,0}_S
    \ket{0,1}_I-\sqrt{\frac{1}{3}}\ket{1,1}_S\ket{1,0}_I ,
\end{equation}
i.e. the searched cloning transformation (\ref{hbgen}). It is
straightforward to see that if the two photons in the signal mode
are separated, for instance by means of a beam-splitter, the
obtained fidelity is equal to 5/6. Indeed the first term
corresponds to ideal cloning, while only one of the two photons in
the second term is equal to the initial state, so
\begin{equation}
    F=1\times
    \frac{2}{3}+\frac{1}{2}\times\frac{1}{3}=\frac{5}{6} .
\end{equation}
The factor $\sqrt 2$ is a manifestation of the stimulated emission
process. It only appears when the initial photon is {\sl
completely} indistinguishable from the down-converted photon in
the signal mode. That is, the two photons should perfectly overlap
in space, time and frequency. Any effect increasing the
distinguishability of these two photons, such as a difference in
the coherence lengths of the pump pulse and down-converted
photons, must be compensated in order to achieve a near to optimal
cloning. Moreover, it has to be stressed that this implementation
of the cloning machine is conditioned on the fact that the three
detectors (the one for the idler mode and the two in the state
analyzers) click. Then, it is assumed that one photon was present
in each mode. Note that there are cases in which more than one
pair is produced by the crystal, or the initial state to be cloned
actually contains more than one photon. These spurious processes
slightly decrease the optimality of the cloning transformation. In
any case, the reported fidelities are equal to $0.81\pm 0.01$
\cite{lam02} and $0.810\pm 0.008$ \cite{dem02,pel03,dem04}, very
close to the theoretical value $5/6\approx 0.83$. Interestingly,
the photon in the idler mode, or anti-clone, gives the optimal
realization of the quantum universal NOT gate. The optimal
fidelity for this transformation is $2/3$, while the reported
experimental fidelity is $0.630\pm 0.008$
\cite{dem02,pel03,dem04}.

\begin{center}
\begin{figure}
\includegraphics[width=8cm]{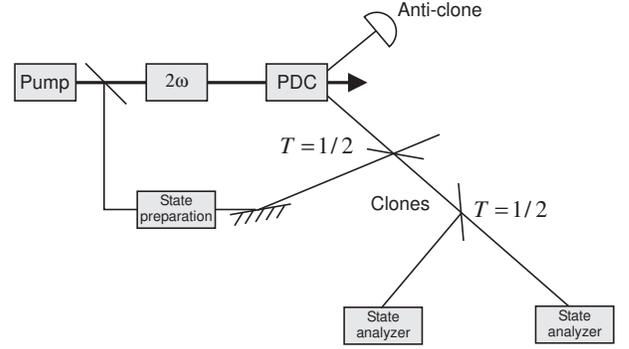}
\caption{Alternative
implementation of the $1\rightarrow 2$ universal cloning machine
for qubits. The photon to be cloned is now combined into a
balanced beam-splitter with one of the down-converted photons.
When the two photons stick together, two optimal clones of the
initial state are obtained. As in the first scheme, the second
down-converted photon provides the anti-clone.} \label{figexpcl2}
\end{figure}
\end{center}

More recently, an alternative version of the $1\rightarrow 2$
quantum cloner for qubits has been proposed and carried out by
\textcite{irv04} and by \textcite{ric04}. This is based on the
fact that two identical photons bunch at a beam-splitter. The
experiment is much simpler but cannot be generalized to
$N\rightarrow M$ cloning. The experimental set-up\footnote{Note
that this is the same set-up as for the teleportation of a qubit
\cite{ben93}.} is schematically shown in Fig.~\ref{figexpcl2}. The
initial state is combined with one of the down-converted photons
into a balanced beam-splitter. It is a well-known result that if
the photons separate after the beam-splitter, a projection onto
the singlet state $\ket{\Psi^-}$ has been achieved. In the other
cases, the photons have been projected with
\begin{equation}
    S_2=\one-\ket{\Psi^-}\bra{\Psi^-} ,
\end{equation}
onto the two-qubit symmetric subspace. Tracing out the second
down-converted photon, the transformation on the photons impinging
the beam-splitter is indeed equal to (\ref{clonwerner}),
conditioned on the fact that they stick together. On the other
hand, it is straightforward to see that the transformation on the
second down-converted photon is the optimal U-NOT gate, i.e. the
photon in the idler mode is equal to the anti-clone (compare with
\ref{ssanti}). In a similar way as for the previous
implementation, the quality of the cloning process crucially
depends on the fact that the two photons arriving at the
beam-splitter define the same mode. This means that, as above,
they have to be completely indistinguishable. Moreover,
multi-photon pulses also deteriorate the quality of the cloning
process. The observed fidelities for cloning were approximately
0.81.

\subsubsection{Proposals for asymmetric cloning}
\label{sssasymexp}

In this section we show how the previous realizations can be
modified in order to cover asymmetric cloning machines. Indeed, it
has been shown very recently that some of these transformations
can be obtained by combining into beam-splitters the photons
produced by a symmetric cloning machine
\cite{fil04,ibl04,ibl05,fiu05}. At the moment of writing, these
experiments have not yet been performed.

\begin{figure}
\includegraphics[width=8cm]{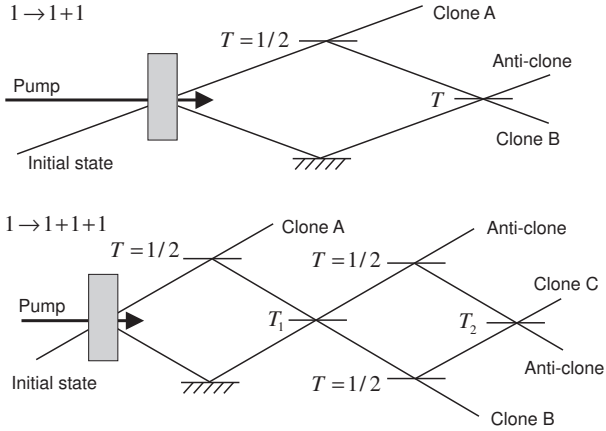}
\caption{The figure shows the optical implementation of the $1\to
1+1$ and $1\to 1+1+1$ optimal cloning machines. For the
$1\rightarrow 1+1$ case, the symmetry between the clones of the
$1\rightarrow 2$ cloning machines is broken by combining one of
the clones with the anti-clone into a beam-splitter. The degree of
asymmetry depends on the transmittivity $T$. The idea can be
naturally generalized to the $1\rightarrow 1+1+1$ case. }
\label{figasexp}
\end{figure}

A proposal for the experimental realization of the asymmetric
$1\rightarrow 1+1$ cloning machine for qubits was given by
\textcite{fil04}. It is represented in Fig.~\ref{figasexp}. It is
convenient for the analysis of this scheme to rewrite the output
of the symmetric machine (\ref{PDCout12}) using Cerf's formalism,
\begin{equation}\label{PDCoutcerf}
    \ket{\psi_{1\rightarrow 2}}=\Big[\frac{\sqrt 3}{2}\one+
    \frac{1}{2\sqrt
    3}\sum_{k=x,y,z}(\sigma_k\otimes\sigma_k\otimes\one)
    \Big]\ket{\psi}\ket{\Psi^-} ,
\end{equation}
where $\ket{\Psi^-}$ is the singlet state. It is simple to see
that this state is equivalent to Cerf's construction when $v=\sqrt
3/2$ (\ref{cerfimplicit}), but with a simple relabelling of the
Bell states for the second clone and the anti-clone. The asymmetry
between the clones can now be introduced by changing the ratio of
the amplitudes for the first term and the rest. A possible way of
achieving this is by successfully applying the projector
$P_{asym}=a\one+b\ket{\Psi^-}\bra{\Psi^-}$ to the second clone and
the anti-clone, where $a$ and $b$ are such that $\one-P_{asym}\geq
0$. Indeed, the states $P_{asym}\ket{\psi_{1\rightarrow 2}}$
define, up to normalization, the family of states
(\ref{cerfimplicit}). Changing the ratio between $a$ and $b$, one
can optimally adjust the asymmetry between the quality of the two
clones, i.e. the ratio between $v$ and $x$ according to Cerf's
notation. A beam-splitter of transmittivity $T$ conditioned on the
fact that the photons at the output are separated gives a simple
optical implementation of this projector. Indeed, some simple
algebra shows that the corresponding operation is equal to
\begin{equation}
    P_{asym}(T)=(2T-1)\one+2(1-T)\ket{\Psi^-}\bra{\Psi^-} .
\end{equation}
When $T=1$ no operation is performed, $P_{asym}(T=1)=\one$, and
the two clones are symmetric, $F_A=F_B=5/6$. When $T$ decreases,
some asymmetry is introduced between the two clones, since $F_A$
increases while $F_B$ worsens. In the limiting case $T=1/2$, a
projection onto the singlet is achieved, as expected, and the
state in mode $A$ is projected onto the initial state, $F_A=1$ and
$F_B=1/2$, that is, the cloning transformation has been undone. In
fact, since a projection onto the singlet is realized, the
detected photons after the beam-splitter were the ones produced in
the crystal. This implies that the photon in mode $A$ must be
equal to the initial state. All the interesting values lie between
these two limiting cases, $1/2\leq T\leq 1$. Indeed, one can see
that the trade-off between the obtained fidelities $F_A$ and $F_B$
saturates the cloning inequality (\ref{clonineq}).

Note that for all these experimental proposals, the successful
implementation of the searched cloning transformation depends on
the detection of three photons (all the detectors click).
Interestingly, one can see that changing the number of
post-selected photons gives other asymmetric cloning machines
\cite{ibl04,ibl05,fiu05}, in a way similar to what happens for the
symmetric case \cite{sim00}. Indeed, denoting by $N$, $M_A$ and
$M_B$ the number of photons in the initial mode and modes $A$ and
$B$ (see Fig.~\ref{figasexp}), it has been shown that the optimal
$1\rightarrow 1+2$ cloning machine is recovered when $N=1$ and
$M_A=1,M_B=2$ and $M_A=2,M_B=1$, and also the $2\rightarrow 2+1$
case when $N=2$ and same post-selection for modes $A$ and $B$.
Unfortunately, the transformation when $N=M_A=M_B=2$ does not
correspond to the optimal $2\rightarrow 2+2$ machine. At present,
it seems that the previous construction only works for the
$1\rightarrow 1+N$ and $N\rightarrow N+1$ cases, and a feasible
optical implementation of the $N\rightarrow M_1+M_2$ machine, with
$M_1,M_2>1$ remains
as an open question. 

Remarkably, Filip's construction can be further generalized.
Indeed, exploiting the anti-symmetrization by means of
beam-splitter allows us to extend this scheme to the $1\rightarrow
1+1+1$ case, where three copies of the initial state are produced
in such a way that the trade-off between the fidelities is
optimal. As shown in Fig.~\ref{figasexp}, it is possible to
consider a more complex situation when the production of two pairs
by the pump pulse, instead of one, is stimulated by the presence
of the photon to be cloned. The corresponding state is equal to
the output of a $1\rightarrow 3$ symmetric machine, as discussed
by \textcite{sim00}. Actually, there are three clones and two
anti-clones, namely the two photon in the idler mode. Now, one can
apply twice the anti-symmetrization explained above, as shown in
Fig.~\ref{figasexp}. After much algebra, one can see that the
fidelities for the clones in modes $A$, $B$ and $C$ are equal to
those defining the optimal $1\rightarrow 1+1+1$ cloning machine of
\textcite{ibl04}, \textcite{ibl05} and \textcite{fiu05}. Although
unproven, it seems quite likely that this construction works for
any number of clones, and that all $1\rightarrow 1+\ldots+1$
cloning machines can be optimally realized by combining into
beam-splitters, and conditioned on the number of photons, the
output of the $1\rightarrow N$ symmetric machine.

\subsubsection{Cloning in an erbium-doped fiber}
\label{sssfiber}

\begin{figure}
\includegraphics[width=8cm]{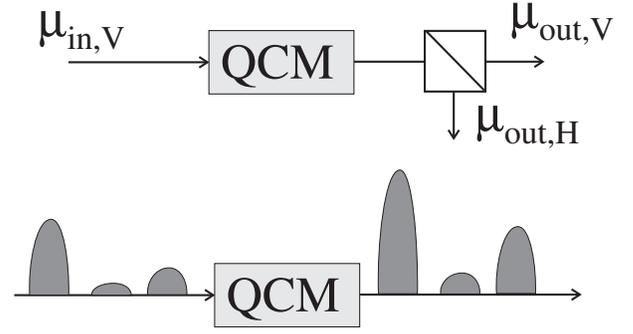}
\caption{Scheme of the experimental set-up of Ref.~\cite{fas02},
see paragraph \ref{sssfiber}. Everything takes place into optical
fibers, the QCM itself being a pumped Er-doped fiber. Upper
figure: effectively realized experiment for cloning of
polarization states (qubit). Lower figure: possible variation
using time-bin encoding, that would lead to optimal cloning of
qudits (here, $d=3$).} \label{figfasel}
\end{figure}

Parametric down-conversion is an amplification medium that has
been studied intensively because it allows to create entangled
photons. In the field of telecommunication optics, however, the
common device used for amplification of light are optical fibers
doped with erbium ions. These rare-earth ions can be pumped onto
an excited state and then constitute an inverted medium that can
lase at telecom wavelengths. \textcite{fas02} studied quantum
cloning due to such an amplifier. The experiment consisted of
sending {\em classical, very weak} pulses of (say) vertically
($V$) polarized light into an erbium-doped fiber. At the output,
light is amplified, but is no longer perfectly polarized because
of spontaneous emission: some light has developed in the
polarization mode orthogonal to the input one (horizontal, $H$).
The fidelity of the classical amplification is defined as the
ratio of the intensities $F_{cl}=I_{out,V}/I_{out, tot}$.

A theoretical analysis based on a seminal paper on maser
amplification \cite{shi57} provides a remarkable prediction: let
$\mu_{in,out}$ be respectively the mean number of photons in the
input and the output field (i.e., the intensity of these fields,
in suitable units). Then it holds \ba
F_{cl}&=&\frac{Q\mu_{out}\mu_{in}+\mu_{out}+
\mu_{in}}{Q\mu_{out}\mu_{in}+2\mu_{out}}\,. \label{clonq}\ea Here,
the parameter $Q$ is related to the phenomenology of the emission
process: $Q=1$ means that all Erbium ions are excited, so that
there is no absorption; $Q=0$ means that emission and absorption
compensate exactly, $Q<0$ means that the absorption in the medium
overcomes the emission. We see that in the ideal case $Q=1$, the
formula (\ref{clonq}) for $F_{cl}$ looks exactly like the one for
the optimal symmetric $N\rightarrow M$ cloning of qubits
(\ref{fqubits}), but for the fact that $\mu_{in}$ and $\mu_{out}$
are not restricted to take integer values. This is a signature of
the underlying quantum cloning in an experiment with classical
states of light. In the actual experiment, the fit yielded
$Q=0.8$; for the cloning $\mu_{in}=1\rightarrow
\mu_{out}=1.94\simeq 2$, a fidelity $F_{cl}\approx 0.82$ was
observed, close to the optimal value $\frac{5}{6}\simeq 0.833$.

Although the experiment was performed with polarization, the same
setup would allow the cloning of quantum states encoded in {\em
time-bins}. With time-bin encoding, it is very easy to go beyond
the qubit case \cite{the04,der20}. In particular, the present
setup (Fig.~\ref{figfasel}) would allow to demonstrate optimal
cloning for higher-dimensional quantum systems. As an example to
support this claim, we compute the fidelity in the computational
basis for $1\rightarrow 2$ cloning --- that is, one photon was
prepared in a given time-bin, and two photons are found in the
outcome. The probability of finding the new photon in the good
time-bin (associated to $F=1$) is just twice the probability of
finding it in any of the other $d-1$ time-bins (in which case
$F=1/2$, because half of the times we pick the original photon).
The average fidelity is then \ba F_{1\rightarrow
2}(d)&=&\frac{2\times 1+(d-1)\times
\frac{1}{2}}{2+(d-1)}\,=\,\frac{d+3}{2(d+1)} \ea which is the
optimal result, see \ref{sswern}. Of course, one should show that
the same fidelity holds for any superposition state, which is,
however, quite evident when one is familiar with the physics of
light amplification. As we mentioned above, this result is not
limited to time-bins, but holds for any encoding of a qudit in
different modes of the field \cite{fan02}; the time-bin encoding
is possibly the most easily analyzed and implemented.

\subsection{Other quantum systems}

\subsubsection{Nuclear spins in Nuclear Magnetic Resonance}

A way to achieve quantum cloning of nuclear spins using Nuclear
Magnetic Resonance (NMR) has been presented by \textcite{cum02},
together with its experimental realization. As usual in quantum
information processing with NMR, many molecules are present in the
sample and the process takes place among nuclear spins within each
molecule.

In the present experiment, the molecule is {\em
E}-(2-chloroethenyl)phosphonic acid. After the peculiar pulse
sequences needed to prepare the sample in a {\em pseudo-pure
state}, a spin direction is encoded into the first qubit, which is
the spin of the $^{31}$P nucleus. The main part of the scheme is a
pulse sequence that implements a version of the optimal symmetric
$1\rightarrow 2$ QCM \cite{buz97} that maps the quantum
information onto the two other qubits --- here, nuclear spins of
two $^1$H atoms. Because of several unwanted mechanisms and
imperfections, however, the measured fidelity for both clones was
only $F\approx 0.58$, even lower than the value achievable with
trivial cloning strategies (see \ref{ssstriv}).

\subsubsection{Atomic states in cavity QED}

Implementations of the $1\rightarrow 2$ UQCM for qubits using the
techniques of cavity QED have been proposed. The scheme by
\textcite{mil03} uses four Rydberg atoms interacting with two
cavities. Atom 2 carries the input state. After the suitable pulse
sequence, Atoms 3 and 4 are the two clones: as in the NMR
experiment described just above, the transformation is similar to
the one of \textcite{buz97}. Here however, the "circuit" is a new
one, and the ancilla is not a single qubit, but two atoms (1 and
2) and the state of the light field in the two cavities.

\textcite{zou03} proposed a scheme that uses three atoms and three
cavities; interaction between atoms within each cavity is required
for this scheme.

\section{Perspectives}
\label{sec7}

\subsection{Some open questions}

At the end of this review, we address a few of the questions that
are still open at the moment of writing. As far as possible, we
list them in the same order as the corresponding themes appear in
this review.

\begin{itemize}

\item To our knowledge, all the study of optimal cloning has
always supposed pure input states. The {\em optimal cloning of
mixed states} is thus a completely open domain. Also, as we
mentioned several times, there is no general result concerning
{\em state-dependent cloning}, and the zoology of cases is a
priori infinite.

\item We have seen several times in this review (especially
Sections \ref{sec3} and \ref{sec4}) that the single-copy fidelity
is not always the most meaningful figure of merit. However, most
of the QCMs are optimal according to it. What about {\em other
figures of merit}? If the resulting QCMs are found to be
different, is there a deep connection among all the results?

\item The role of entanglement in cloning may be further
elucidated. The trivial strategies discussed in paragraphs
\ref{ssstriv} and \ref{ssstr2} show that the fidelity $F_{triv}$
given in Eq.~(\ref{ftriv}) can be achieved without using any
coherent interaction between the clone and the copy. It seems
quite plausible that this is the optimal value one can attain
using strategies without quantum interaction. Therefore, it would
be interesting to understand more precisely what role entanglement
plays in optimal cloning, e.g. by studying the entangling power of
the optimal cloning machine or the entanglement properties of the
corresponding output states \cite{bru03}. Note that in the limit
of large dimension, no entanglement is required for an optimal
cloning. One could also look for links between these results and
the entanglement cloning machine of \textcite{lam04}.

\item Also the link between {\em cloning and Bell's inequalities}
is not clear. Consider the setup of Fig.~\ref{figsignaling}. The
QCM is an existing one (not Herbert's hypothetical perfect
cloner), and let's suppose it universal and symmetric for
simplicity. Alice and Bob started with the singlet, which
obviously violates Bell's inequalities. Alice keeps the quantum
system A, Bob now has two quantum systems B$_1$ and B$_2$. Does
$\rho_{AB_1}=\rho_{AB_2}$ violate a Bell inequality? Certainly, it
cannot violate any inequality with two settings on Bob's side,
because Bob could measure one setting on $\rho_{AB_1}$ and the
other setting on $\rho_{AB_2}$ \cite{ter03}. But the general
answer is unknown.

\item The {\em connection between optimal cloning and state
estimation} looks natural and, indeed, in paragraph \ref{ss4meas}
we presented several results in that direction. However, it is
still not known whether this connection holds in general. Is it
true for any arbitrary set of states, possibly with unequal
a-priori probabilities, that the fidelities are equal for the
optimal state estimation and for the optimal cloning in the limit
of a large number of copies?

\item The {\em relation to optimal eavesdropping} is also not yet
fully understood. For individual attacks on some quantum
cryptography protocol, such as BB84 or the six-state protocol, it
has been proven that the best strategy uses the cloning machines
that are optimal to clone the set of states used for encoding. As
we stressed in Section \ref{sec4}, this correspondence is not
obvious, since cloning is optimized for fidelities, whereas in
eavesdropping one optimizes mutual Shannon information; and
indeed, it seems that the correspondence breaks down for the
SARG04 protocol. More generally, it has been proven that security
bounds can be obtained by restricting attacks to the so-called
"collective attacks" \cite{kra04,ren05}, and it is meaningful to
ask whether the quantum interaction is described by the
corresponding optimal cloner.

\item The concepts and tools of cloning have proved useful for the
foundations of quantum mechanics, for state estimation and for
cryptography. Are there other domains, tasks, situations, etc. in
which cloning can be useful? Or, can one find a more general
principle which unifies optimal cloning, state estimation, and
eavesdropping in cryptography, possibly with spontaneous and
stimulated emission?

\item Qubits obey fermionic commutation relations, e.g.
$\{\sigma_x,\sigma_y\}  = 0$. Optimal cloning of qubits can be
implemented using spontaneous and stimulated emission that comes
from bosonic commutation relations, ie $[a,a^{\dagger}]=1$. What
is the exact relation? A link between the particle statistics and
state estimation has been discussed \cite{bos03}, but to our
knowledge there is no such study for cloning.

\item Many questions are also still open in the field of {\em
implementations} of quantum cloning. Obviously, any form of
cloning can (in principle) be implemented with linear optics using
the \textcite{kni01} scheme for quantum computation. Can one
implement any cloning transformation using amplification through
stimulated emission? If yes, can one it be done by linear-optics
elements, or are other non-linear devices needed? Are there other
"natural" phenomena that directly implement quantum cloning?

\end{itemize}

This list will possibly shrink in the coming years as soon as
these questions are answered. A regularly updated list of open
problems in quantum information is available on the website of
Reinhard Werner's group:
www.imaph.tu-bs.de/qi/problems/problems.html. At the moment of
this writing, no problems related to cloning are listed there,
apart from, possibly, "Complexity of product preparations"
proposed by Knill.

\subsection{Conclusion: the role of cloning in quantum physics}

Quantum cloning is likely to remain an active topic for basic
research, while simultaneously an ideal subject for teaching
elementary quantum physics. The proof of the no-cloning theorem is
so simple that it can be presented to students as soon as the
linearity of the quantum dynamics has been introduced, and much of
quantum mechanics can be presented as a consequence of this deep
no-go theorem. Such a presentation would not follow the history of
the discovery of quantum physics, but is much closer to the modern
view of it in the light of quantum information theory. Optimal
cloning clearly shows that incompatible quantities can be measured
simultaneously (first clone the system, next perform different
measurements on each clone), while illustrating that such
measurements can't be ideal, i.e. can't be immediately
reproducible.

Apart from the issue of measurement, quantum cloning is closely
related to many other aspects of quantum physics: to the
no-signaling condition, both historically (\ref{sechist}) and as
limit for optimal cloning (\ref{secsign}); to the phenomenon of
spontaneous and stimulated emissions, well known in quantum
optics, see Section \ref{sec5}... The no-cloning theorem also
introduces in a natural way the idea of quantum cryptography, and
optimal cloning suggests the way eavesdropping can be analyzed.
Finally, it elucidates what is so special about quantum
teleportation \cite{ben93}: the original has to be destroyed in
the process and the Bell-state measurement should not provide any
information of the state to be teleported, otherwise there would
be a contradiction with the no-cloning theorem (more precisely,
with the optimal asymmetric cloning result presented in
\ref{ssasym}).

\section{Acknowledgements}

V.S. acknowledges the invitation of the {\em Troisi\`eme cycle de
la physique en Suisse romande} to lecture in the course "Quantum
communication", as these lectures formed the starting point of the
present review. We acknowledge financial support from the European
project RESQ, the Swiss NCCR "Quantum photonics" (V.S., N.G.,
S.I.), the Spanish MCYT under the "Ram\'on y Cajal" grant and the
Generalitat de Catalunya (A.A.).

\begin{appendix}

\section{Notations and basic formulas for qubits}
\label{appa}

A {\em qubit} is the simplest possible quantum system, described
by the two-dimensional Hilbert space $\compl^2$. The algebra of
operators acting on this space is generated by the Pauli matrices:
\ban \si_x=\left(\begin{array}{cc} 0 & 1\\ 1 &
0\end{array}\right),&\, \si_y=\left(\begin{array}{cc} 0 & -i\\
i & 0\end{array}\right),&\, \si_z=\left(\begin{array}{cc} 1 & 0\\
0 & -1\end{array}\right)\,.\ean In particular,
$\mbox{Tr}(\si_k)=0$ and $\si_k^2=\one$ for $k=x,y,z$; also,
$\si_x\si_y=-\si_y\si_x = i\si_z$ and all the cyclic permutations
hold. In the set of states, the computational basis
$\big\{\ket{0},\ket{1}\big\}$ is universally assumed to be the
eigenbasis of $\si_z$, so that: \ba
\si_z\ket{0}\,=\,\ket{0}\,&,&\, \si_z\ket{1}\,=\,-\ket{1}\,. \ea
Normally, everything is always written in the computational basis;
only the eigenstates of $\si_x$ have a standard notation for
convenience: $\ket{+x}\equiv\ket{+}\,=\,
\frac{1}{\sqrt{2}}\big(\ket{0} + \ket{1}\big)$ and
$\ket{-x}\equiv\ket{-} \,=\,\frac{1}{\sqrt{2}}\big(\ket{0} -
\ket{1}\big)$. The eigenstates of $\si_y$ are $\ket{\pm y}\,=\,
\frac{1}{\sqrt{2}}\big(\ket{0} \pm i\ket{1}\big)$.

The generic {\em pure state} of a qubit will be written \ba
\ket{\psi}&=&\alpha\ket{0}\,+\,\beta\ket{1}\,. \ea The associated
projector reads \ba \ket{\psi}\bra{\psi}&= &\demi\, \Big(\one\,
+\, \hat{n}\cdot\vec{\si} \Big) \label{proj1}\ea where the vector
$\hat{n}\,=\,(\moy{\si_x}_{\psi},
\moy{\si_y}_{\psi},\moy{\si_z}_{\psi}) =
(2\mbox{Re}(\alpha\beta^*),2\mbox{Im}(\alpha\beta^*),|\alpha|^2-
|\beta|^2)$ is called Bloch vector. For pure states (the case we
are considering here), its norm is 1: actually, all these vectors
cover the unit sphere (called the {\em Bloch sphere}, or the
Poincar\'e sphere if the two-level system is the polarization of
light). Thus, there is a one-to-one correspondence between unit
vectors and pure states of a two-level system given by the
following parametrization in spherical coordinates: \ba
\ket{\psi}\,\equiv\,\ket{+\hat{n}}\,=\,\cos\frac{\theta}{2}\,\ket{0}\,+\,
e^{i\varphi}\sin\frac{\theta}{2}\,\ket{1}\ea is the eigenstate for
the eigenvalue $+1$ of $\hat{n}\cdot\vec{\si}$, with
$\hat{n}\equiv \hat{n}(\theta,\varphi) =(\sin\theta\cos\varphi,
\sin\theta\sin\varphi, \cos\theta)$, with as usual
$\theta\in[0,\pi]$ and $\varphi\in[0,2\pi]$. Given that any
projector takes the form (\ref{proj1}), the general form of any
mixed state can then be written: \ba \rho\,=\,\sum_k
p_k\ket{\psi_k}\bra{\psi_k}&=& \demi\, \Big(\one\, +\,
\vec{m}\cdot\vec{\si} \Big)\ea with $\vec{m}=\sum_k p_k\vec{m}_k$;
the norm of the Bloch vector is $|\vec{m}|\leq 1$, with equality
if and only if the state is pure.

For the present review, it is also useful to mention some formulae
and notations for the description of {\em two qubits}. As is
well-known, a composed system is described by the tensor product
of the Hilbert spaces of its components. So the Hilbert space that
describes a two-qubit system is ${\cal
H}=\compl^2\otimes\compl^2$. The natural ("induced") computational
basis on this space is the basis of the four eigenstates of
$\si_z\otimes\si_z$, namely (we omit the symbol of tensor product
for states, when not necessary) $\ket{0}\ket{0}$,
$\ket{0}\ket{1}$, $\ket{1}\ket{0}$ and $\ket{1}\ket{1}$. The most
general pure state is any linear combination of these. Although
probably redundant in a paper on quantum information, we recall
here that the most important feature of composed systems is the
existence of {\em entangled states}, that is, states that cannot
be written as products $\ket{\psi_1}\otimes\ket{\psi_2}$.

The basis formed with four orthogonal maximally entangled states
({\em Bell basis}) plays an important role; the notations are
standardized by now: \ba
\ket{\Phi^+}&=&\frac{1}{\sqrt{2}}\,\big(\ket{0}\ket{0} +
\ket{1}\ket{1} \big)\,,\\
\ket{\Phi^-}&=&\frac{1}{\sqrt{2}}\,\big(\ket{0}\ket{0} -
\ket{1}\ket{1} \big)\,,\\
\ket{\Psi^+}&=&\frac{1}{\sqrt{2}}\,\big(\ket{0}\ket{1} +
\ket{1}\ket{0} \big)\,,\\
\ket{\Psi^-}&=&\frac{1}{\sqrt{2}}\,\big(\ket{0}\ket{1} -
\ket{1}\ket{0} \big)\,.\ea We recall that $\ket{\Psi^-}$ is
invariant under identical unitaries on both qubits, i.e. it keeps
the same form in all the bases. If the eigenstates of
$\si_x\otimes\si_x$ or of $\si_y\otimes\si_y$ were taken as
computational bases states, the Bell basis remains the same,
simply relabelled: $\ket{\Phi^{+}}_x
=\ket{\Psi^{+}}_y=\ket{\Phi^{+}}_z$,
$\ket{\Phi^{-}}_x=-i\ket{\Phi^{-}}_y=\ket{\Psi^{+}}_z$ and
$\ket{\Psi^{+}}_x=\ket{\Phi^{+}}_y=\ket{\Phi^{-}}_z$.

The general form of a density matrix of two qubits is \ba
\rho_{AB}&=&\frac{1}{4}\,\big(\one_4\,+\,\hat{n}_A\cdot\vec{\si}\otimes\one
\,+\, \one\otimes \hat{n}_B\cdot\vec{\si}\nonumber\\
&&+\,\sum_{i,j=x,y,z}\, t_{ij}\si_i\otimes \si_j \big) \ea where
$t_{ij}=\mbox{Tr}(\rho\,\si_i\otimes \si_j)$. From this form, the
partial traces are computed leading to \ba \rho_{A,B}&=&\demi\,
\left(\one\,+\,\hat{n}_{A,B}\cdot\vec{\si}\right)\,.
\label{mix2}\ea If $|\hat{n}_A|=1$, then $\rho_A=P_A$ a projector
on a pure state; since a projector is an extremal point of a
convex set, this necessarily implies $\rho_{AB}=P_A\otimes \rho_B$
and in particular $t_{ij}=(n_A)_i\,(n_B)_j$.

\end{appendix}

\bibliographystyle{apsrmp}


\end{document}